\documentclass{article}

\usepackage{arxiv}

\usepackage[utf8]{inputenc} 
\usepackage[T1]{fontenc}    
\usepackage{url}            
\usepackage{booktabs}       
\usepackage{amsfonts}       
\usepackage{nicefrac}       
\usepackage{microtype}      
\usepackage{graphicx}
\usepackage{doi}
\usepackage[all]{xy}
\usepackage{setspace}
\usepackage{xcolor}
\usepackage{tikz}
\usetikzlibrary{cd}

\title{On the role of geometry in statistical mechanics and thermodynamics I: Geometric perspective}

\author{O{\u g}ul Esen\\
Department of Mathematics, Gebze Technical University,\\
41400 Gebze, Kocaeli, Turkey\\
Corresponding author: oesen@gtu.edu.tr\\
\And Miroslav Grmela\\
\'{E}cole Polytechnique de Montr\'{e}al,
  C.P.6079 suc. Centre-ville,\\
 Montr\'{e}al, H3C 3A7,  Qu\'{e}bec, Canada\\
\And and Michal Pavelka \\
 Mathematical Institute, Faculty of Mathematics, Charles University,\\ 
 Sokolovsk\'{a} 83, 18675 Prague, Czech Republic
}

\newcommand{\shorttitle}{Geometric perspective}

\hypersetup{
pdftitle={On the role of geometry in statistical mechanics and thermodynamics I: Geometric perspective},
pdfsubject={}
pdfauthor={O. Esen, M. Grmela, and M. Pavelka},
pdfkeywords={Geometric lifts, statistical mechanics, Hamiltonian mechanics}
}

\counterwithin*{equation}{section} 
\usepackage{amsmath,amssymb,amsfonts,latexsym,float,graphics,epsfig}

\newtheorem{theorem}{Theorem}[section]
\newtheorem{proposition}[theorem]{Proposition}

\newcommand{\p}{\partial}

\newcommand{\qq}{\mathbf{q}}
\newcommand{\pp}{\mathbf{p}}

\newcommand{\xx}{\mathbf{x}}

\begin{document}
\maketitle

\begin{abstract}
This paper contains a fully geometric formulation of the General Equation for Non-Equilibrium Reversible-Irreversible Coupling (GENERIC).
Although GENERIC, which is the sum of Hamiltonian mechanics and gradient dynamics, is a framework unifying a vast range of models in non-equilibrium thermodynamics, it has unclear geometric structure, due to the diverse geometric origins of Hamiltonian mechanics and gradient dynamics. 
The difference can be overcome by cotangent lifts of the dynamics, which leads, for instance, to a Hamiltonian form of gradient dynamics.
Moreover, the lifted vector fields can be split into their holonomic and vertical representatives, which provides a geometric method of dynamic reduction. The lifted dynamics can be also given physical meaning, here called the rate-GENERIC. Finally, the lifts can be formulated within contact geometry, where the second law of thermodynamics is explicitly contained within the evolution equations.
\end{abstract}

\tableofcontents

\doublespace

\section{Introduction}
An objective of statistical mechanics and thermodynamics is to transform mechanics of $\sim 10^{23}$ particles into various reduced forms (like hydrodynamics or equilibrium thermodynamics), that emerge when observing macroscopic systems. For instance, hydrodynamics is an autonomous theory that has been introduced by Euler \cite{euler} without any reference to the fact that fluids are composed of a large number of particles obeying Newton's mechanics. Similarly, the classical equilibrium thermodynamics combining Newton's mechanics of macroscopic bodies with heat arose from direct macroscopic observations and not from seeing macroscopic systems as composed of microscopic particles. The passage from microscopic to macroscopic descriptions, if established, brings understanding of various new features emerging in macroscopic physics. That, in turn, allows to control the behaviour of macroscopic systems by tweaking their microscopic constituents (the goal of material science and nanotechnologies).

The transformation  of  microscopic dynamics into its macroscopic  form follows three steps. The first step consists of finding the microscopic phase portrait (collection of all microscopic particle trajectories). A pattern is then identified in the microscopic phase portrait in the second step. In the third step, the pattern is interpreted as a phase portrait corresponding to the macroscopic dynamics. All three steps are obviously very difficult to make, and any method or tool that can be useful is welcome.

The name statistical mechanics was coined by Maxwell and the use of stochastic structures in the pattern recognition is indeed very natural and useful. Nevertheless, the structures based in geometry have also been found helpful (for instance in the widely used face recognition algorithms) and geometric methods have indeed been used and contributed substantially to statistical mechanics and thermodynamics \cite{callen,Wolf2009}. For example, casting particle mechanics into the form of the Liouville equation (the first step in the standard approach to statistical mechanics) is in fact a probabilistic reinterpretation of the original Liouville' work \cite{Liouville}, relating a system of ordinary differential equations to a partial differential equation. The dynamics taking place on the microscopic state space (described by ordinary differential equations) is lifted to dynamics of sections of a bundle with the microscopic state space as its base space. This original viewpoint of the Liouville equation has been then followed in 
\cite{carleman,Koopman}. Another example of the use of geometry in statistical mechanics and thermodynamics is the original Gibbs formulation of the classical equilibrium thermodynamics \cite{gibbscw}. The geometrical viewpoint of thermodynamics has been then followed in several directions, in particular the contact-geometric formulation of equilibrium thermodynamics \cite{Hermann,Mrugala} and non-equilibrium thermodynamics in \cite{grmela2010advchemeng}.

A wide range of phenomena in non-equilibrium thermodynamics can be formulated within the framework of General Equation for Non-Equilibrium Reversible-Irreversible Coupling (GENERIC) \cite{go,og}, including fluid mechanics, kinetic theory, dynamics of complex fluids, electrodynamics, solid mechanics, or dynamics of mixtures \cite{hco,pkg}. Within GENERIC, the evolution of a system (described by state variables $\xx$) is the sum of Hamiltonian mechanics and generalized gradient dynamics, 
\begin{equation}
\dot{\xx}=\{\xx,E\} + \Xi_{\xx^*}|_{\xx^*=S_\xx},
\end{equation}
where $E$ is the energy of the system, $S$ its entropy, $\{\bullet,\bullet\}$ is a Poisson bracket, and $\Xi(\xx,\xx^*)$ a dissipation potential. Despite the wide applicability, this framework has a fundamental inconsistency. Although both Hamiltonian mechanics and gradient dynamics are well-defined geometric concepts separately, what kind of geometry do they form together? This caveat can be illustrated for instance on the Jacobi identity, which is a property of the Poisson bracket ensuring that the Poisson bivector is invariant with respect to the Hamiltonian flow \cite{GF}. However, there is no such property required for the dissipation potential. Is there any geometric consistency behind the GENERIC framework? In this paper, we answer these questions by providing several geometric frameworks suitable for GENERIC.

Our intention in this series of two papers is to explore another geometrical formulation of statistical mechanics and thermodynamics, using lifts in appropriately constructed bundles. The role of the geometric lifts is to cast dynamical systems into forms in which patterns in their phase portraits become easier to recognize. In particular, it allows to lift gradient dynamics to a Hamiltonian form equivalent with the original dynamics if a Hamilton-Jacobi condition is satisfied. The splitting of the lifts to holonomic and evolutionary parts leads to a geometric reduction method in non-equilibrium thermodynamics. Also the lifts in the context of the Morse family turns out to be a geometric way towards constrained dynamics.
In the second paper \cite{OMM-2} we put the geometrical structures introduced in this  paper into the context of equilibrium and nonequilibrium multiscale thermodynamics.

The geometrical presentation in this first paper begins with investigations of the cotangent lifts of the Hamiltonian dynamics and associated with it geometrical Hamilton-Jacobi theory relating the lifted dynamics to its base. Section \ref{2} contains an introduction to Hamiltonian mechanics, complete cotangent lifts, Morse family, and holonomic and vertical (evolutionary) representatives on jet bundles. Hamilton-Jacobi theory then provides a compatibility condition between the original and lifted vector fields. If the condition is violated, setting the vertical representative to zero, we obtain a geometric reduction method from more detailed dynamics (particle with inertia and friction) to a less detailed one (overdamped particle motion).

In Section \ref{3}, we demonstrate that the cotangent and holonomic lifts provide a geometrical setting combining Hamiltonian mechanics with generalized gradient dynamics (generated by a dissipation potential). In particular, the lifts lead to a fully geometric formulation of the GENERIC framework, where both Hamiltonian mechanics and gradient dynamics become part of a single geometric construction, despite their different geometrical origins (skew-symmetric Hamiltonian mechanics and symmetric Riemannian metric). Note that also the metriplectic systems, which can be seen as GENERIC with quadratic dissipation potentials, can be geometrized this way \cite{mor}. As particular examples, we show a Hamiltonian form of chemical kinetics and a geometric formulation of constrained gradient dynamics.
We do not only arrive at GENERIC, but also at its extension (called \textit{rate GENERIC}) that, from the physical point of view,  addresses the stage in the time evolution of macroscopic systems that precedes that stage described by GENERIC. The physical content, the placement in thermodynamics and statistical mechanics, and specific illustrations of both GENERIC and rate GENERIC are discussed in the second paper \cite{OMM-2} of this series. 

In Section \ref{4} we adapt the investigation presented in Section \ref{3} to the contact geometry setting, comparing the two geometrically distinguished vector fields: the canonical contact vector field and the evolutionary vector field. This extension brings to GENERIC the explicit evolution of the thermodynamic potential, which expresses the second law of thermodynamics.

\section{On Geometry of Reversible Motion}\label{2}
This Section recalls Hamiltonian mechanics, the Hamilton-Jacobi theory (HJ), cotangent lifts, holonomic and vertical representatives, and the Morse families. In particular, HJ serves as a compatibility condition making the dynamics before and after lifting equivalent. If the HJ condition is satisfied, then even gradient dynamics, which typically describes irreversible evolution, can be seen as a Hamiltonian system. 

\subsection{Hamiltonian Dynamics}
In this subsection, we recall some basics of Hamiltonian dynamics on symplectic and Poisson manifolds.

\textbf{Symplectic Manifolds.}
Let us start with a configuration manifold $M$ with local coordinates $\mathbf{x}=(x^i)$. Its cotangent bundle $T^{*}M$ admits the canonical (Liouville) one-form $\theta_M$ and the  
symplectic two-form $\Omega_M$, see, for example,  \cite{MR515141,leon89,holm11,Liber87}. The value  of $\theta_M$ on a vector field $X$ over $T^{*}M$ is defined as
		\begin{equation} \label{can-Lio}
		\theta_M(X)= \langle \tau_{T^{*}M}(X), T\pi_M(X)\rangle,
		\end{equation}
	where $\tau_{T^{*}M}$ is the projection from the tangent bundle $TT^{*}M$ to its base manifold $T^{*}M$, whereas $T\pi_M$ is the tangent lift of the cotangent projection $\pi_{M}$. The following commutative diagram summarizes the relations between $M$, its tangent bundle $TM$, cotangent bundle $T^*M$, and $TT^*M$:
			\begin{equation}\label{one-form}
		\begin{tikzcd}
		&TT^*M\arrow[dr,"\tau_{T^*M}"]\arrow[dl,"T\pi_M",swap] \\
		TM \arrow[dr,"\tau_{M}",swap] && T^*M\arrow[dl,"\pi_{M}"] \\
		&M
		\end{tikzcd}
		\end{equation}

	Minus of the exterior derivative of  the canonical one-form $\theta_M$, that is, $\Omega_M:=-d\theta_M$, is the canonical symplectic two-form on the cotangent bundle $T^{*}M$.  In the Darboux' coordinates $(\mathbf{x},\mathbf{x}^*)=(x^i,x^*_i)$ on $T^{*}M$, the canonical forms become
\begin{equation}
\theta_M =\mathbf{x}^* \cdot d \mathbf{x} ,\qquad \Omega_M=d \mathbf{x} \wedge d\mathbf{x}^*.
\end{equation} 
Finally, a submanifold of a symplectic manifold is said to be Lagrangian if (i) the dimension of the submanifold is the half of the dimension of the symplectic manifold and (ii) the symplectic two-form vanishes on that submanifold.

\textbf{Hamiltonian Dynamics on Symplectic Manifolds.}
Each real-valued (Hamiltonian) function on the symplectic manifold $(T^*M,\Omega_M)$ determines a Hamiltonian vector field $X_H$ by
\begin{equation}\label{Ham-Eq}
\iota_{X_H}\Omega_M=dH. 
\end{equation}
In Darboux' coordinates, the Hamiltonian vector field $X_H$ becomes
\begin{equation}  \label{HamEqLoc}
X_{H} (\mathbf{x},\mathbf{x}^*)= H_{\mathbf{x}^*} \cdot \nabla_\mathbf{x} - H_{\mathbf{x}} \cdot \nabla_{\mathbf{x}^*},
\end{equation}
where subindexes indicate partial derivatives (for instance $H_\mathbf{x}$) and $\nabla_\mathbf{x}$ is the vectorial form of the basis of the tangent space, $
\nabla_\mathbf{u}=(\partial_{u^1} ,\dots,  \partial _{u^n})$. 
In this notation, the Hamilton's equations, which represent evolution along the Hamiltonian vector field,  are 
\begin{equation}  \label{HamEqLoc-}
\frac{d \mathbf{x}}{dt} =H_{\mathbf{x}^*} ,\qquad 
\frac{d \mathbf{x}^*}{dt} =-H_{\mathbf{x}}.
\end{equation}

\textbf{Poisson Manifolds.} 
 For a manifold $M$, bracket 
\begin{equation}
\{\bullet, \bullet \} :\mathcal{F}(M)\times \mathcal{F}(M)\longrightarrow \mathcal{F}(M),
\end{equation}
which is defined on the smooth functions $\mathcal{F}(M)$, is called a Poisson bracket when it is satisfying the following properties  \cite{Laurent13,vaisman2012lectures,We83}:
\begin{subequations}\label{propPB}
\begin{align}
& \text{Skew-symmetry:} &\{A,B\}=-\{B,A\},\\
& \text{Bilinearity:} &\{rA+sB,C\}=r\{A,C\}+s\{B,C\},\\
& \text{Jacobi identity:} &\{A,\{B,C\}\}+\{B,\{C,A\}\}+\{C,\{A,B\}\}=0,\\
& \text{Leibniz rule:} &\{A,BC\}=\{A,B\}C+\{A,C\}B,
\end{align} 
\end{subequations}
for all $A$, $B$ and $C$ in $\mathcal{F}(M)$ and for all real numbers $r$ and $s$ in $\mathbb{R}$. The first three requirements in \eqref{propPB} makes $\mathcal{F}(M)$ a Lie algebra. 

A manifold equipped with a Poisson bracket is called a Poisson manifold, and a function  $C$ is called a Casimir function if it commutes with all other functions, $\{A,C\}=0$  for all $A$. If there is no non-constant Casimir function for a Poisson bracket, then we say that the Poisson bracket is non-degenerate, and the Poisson manifold is then also symplectic.

\textbf{Poisson Bivector.}
Once we have a Poisson bracket, we obtain the Poisson bivector field $\mathbb{L}$ as
\begin{equation} \label{bivec-PoissonBra}
\mathbb{L} (dA,dB):=\{A,B\}
\end{equation}
for all $A$ and $B$, where $dA$ and $dB$ denote the exterior (de Rham) derivatives (or in general functional derivatives \cite{hco}). 
In other words, the Poisson bivector can be also represented by a set of coefficient functions $\mathbb{L}=(L^{ij})$ such that
\begin{equation} \label{PB-local}
\{A,B\}=A_{\mathbf{x}} \cdot \mathbb{L} B_{\mathbf{x}} . 
\end{equation}

\textbf{Hamiltonian Dynamics on Poisson Manifolds.}
The Hamilton's equation generated by a Hamiltonian function(al) $E$ on a Poisson manifold reads
\begin{equation}
\frac{d \mathbf{x}}{dt} =\{\mathbf{x},E\}.
\end{equation}
The equation actually expresses evolution along the Hamiltonian vector field $X_{E}$,
\begin{equation}
X_{E}(A)=\{A,E\}, 
\end{equation}
where $X_{E}(A)$ stands for the directional derivative of $A$ in the direction of $X_E$. In particular, the Hamiltonian vector field generated by a Casimir function(al) $C$ is identically zero, which means that $X_E=X_{E+C}$ for any Casimir function(al) $C$. In the finite-dimensional case and local coordinates $\mathbf{x}$ on $M$, the equation of motion generated by a Hamiltonian function $E$ becomes
\begin{equation}\label{Poisson-flow}
X_{E}=\mathbb{L}  E_{\mathbf{x}} \cdot \nabla_{\mathbf{x}},\qquad 
\frac{d \mathbf{x}}{dt}
 =\mathbb{L}  E_{\mathbf{x}} .
\end{equation} 

The following section focuses on complete cotangent lifts of Hamiltonian vector fields, which is the key tool throughout this paper.

\subsection{Lifting Poisson Flow to a Symplectic Flow}\label{Sec-Poisson}

This section contains a geometric method lifting Poisson flows on a manifold $M$ to symplectic flows on the cotangent bundle $T^*M$.  

\textbf{Complete Cotangent Lift.}
Let $M$ be a manifold and $T^*M$ be its cotangent bundle. The complete cotangent lift of a flow $\varphi _{t}$ (of a vector field $X$) on $M$ is
a one-parameter group of diffeomorphisms $\varphi _{t}^{c\ast }$ on $T^{\ast }%
M$ satisfying%
\begin{equation}
\pi _{M}\circ \varphi _{t}^{c\ast }=\varphi _{t}\circ \pi _{%
M},  \label{cotanlift}
\end{equation}%
where $\pi _{M}$ is the canonical cotangent bundle natural projection from $T^{\ast }M
$ to $M$.  
The vector field $%
X^{c\ast }$ on $T^*M$, which has the flow $\varphi _{t}^{c\ast },$ is called the 
complete cotangent lift of $X$, \cite{YaPa67}. The infinitesimal
version of Equation (\ref{cotanlift}) determines $X^{c\ast }$ as 
\begin{equation}
T\pi _{M}\circ X^{c\ast }=X\circ \pi _{M},
\end{equation}
where $T\pi _{M}$ is the tangent mapping of $\pi_M$, which is summarized in the following diagram:
\begin{equation*}
\xymatrix{T^*M \ar[rr]^{X^{c*}}\ar[dd]^{\pi_M}&& TT^*M    \ar[dd]^{T\pi_M}\\ \\  M  \ar[rr]_{X} && TM}
\end{equation*}
The complete cotangent lift of $X$  expresses how both position on the manifold $M$ and covectors (one-forms) vary along the motion induced by the field. 
An important feature of the complete cotangent lift is that taking a vector field to its complete cotangent lift 
is a Lie algebra homomorphism, see for example \cite{EsGu12,Marsden1999}. 

Assume a local coordinate system $\mathbf{x}$ on $M$. A vector field is given as $X(\mathbf{x})=\mathbf{X}(\mathbf{x})\cdot \nabla _{\mathbf{x}}$ and the complete cotangent lift $X^{c*}$ becomes
\begin{equation}
X^{c\ast }=\mathbf{X}\cdot \nabla _{\mathbf{x}}-
\nabla _{\mathbf{x}}(\mathbf{x}^{\ast }\cdot \mathbf{X})\cdot \nabla _{\mathbf{x}^*}.
\label{eq.CCL}
\end{equation}
This vector field is Hamiltonian with respect to the canonical symplectic two-form $d\mathbf{x} \wedge d\mathbf{x}^*$, and it corresponds to the Hamiltonian function $F ( \mathbf{x},\mathbf{x}^{\ast
} ) = \mathbf{x}^*\cdot \mathbf{X}(\mathbf{x})$.

\textbf{The Complete Cotangent Lift of Poisson Flow.} In this case, we consider a Poisson manifold $(M,\mathbb{L} )$ and a Hamiltonian dynamics $X_E$ generated by a Hamiltonian function $E$.  
In the Darboux' coordinates  $\left( \mathbf{x},\mathbf{x}^{\ast }\right) $ the
complete cotangent lift of the Hamiltonian vector field $X_E$ in \eqref{Poisson-flow} is 
\begin{equation} \label{X-c*}
X_E^{c*}(\mathbf{x},\mathbf{x}^*)  = \mathbb{L}  E_{\mathbf{x}} \cdot \nabla_{\mathbf{x}}- \nabla_{\mathbf{x}}\big(
\mathbf{x}^{\ast }\cdot \mathbb{L}   E_\mathbf{x} 
\big)\cdot \nabla_{\mathbf{x}^*}.
\end{equation}
Note that $X_E^{c*}$ is also a Hamiltonian vector field on the cotangent bundle $T^*M$ equipped with the canonical symplectic two-form and corresponds to the  Hamiltonian function
\begin{equation}  \label{F}
F(\mathbf{x},\mathbf{x}^{\ast })=\mathbf{x}^{\ast }\cdot \mathbb{L}   E_\mathbf{x}.
\end{equation}%

Finally, the dynamics determined by the lifted vector field on the cotangent bundle becomes
\begin{equation}
\frac{d \mathbf{x}}{dt}  =\mathbb{L}  E_{\mathbf{x}},\qquad 
\frac{d \mathbf{x}^*}{dt}  =
- \nabla_{\mathbf{x}}\big(
\mathbf{x}^{\ast }\cdot \mathbb{L}   E_\mathbf{x} 
\big),
\end{equation}
where the first set of equations is precisely the Poisson flow \eqref{Poisson-flow}. 

 \subsection{Hamilton-Jacobi Theory}
 \label{sec.HJ}
In this section we recall the Hamilton-Jacobi (HJ) theory, and in particular its stationary version, which will play the role of compatibility condition between the original and lifted dynamics. 

Consider a Hamiltonian dynamics. A Hamilton-Jacobi equation is a partial differential equation for a generating function $W=W(\mathbf{x},t)$, called Hamilton's principal function, on the extended configuration space $M\times \mathbb{R}$ given by
\begin{equation}\label{tdepHJ}
W_t+H\left(\mathbf{x},  W_{\mathbf{x}}\right)=0,
\end{equation}
see, for example, \cite{Arnold-book,Goldstein-book}.

The Hamilton-Jacobi equation \eqref{tdepHJ} and the generating function $W$ are related with the following variational problem. Consider a non-degenerate time-dependent Hamiltonian function $H$ on $T^*M\times \mathbb{R}$ equipped with the Darboux' coordinates $(\mathbf{x},\mathbf{x}^*,t)$. Here, $t$ is the coordinate on the extension which physically refers to the time. Then the (inverse) Legendre transformation of $H$ determines a Lagrangian function  $L=L(\mathbf{x},\dot{\mathbf{x}},t)$ on the extended  tangent bundle $TM\times \mathbb{R}$ with induced coordinates $(\mathbf{x},\dot{\mathbf{x}},t)$. 
Consider the action
\begin{equation}
    W= \int_{t_0}^{t_1} L(\mathbf{x},\dot{\mathbf{x}},t) ~dt .
\end{equation}
Variation of this action evaluated at the extremal trajectories (solutions of the Euler-Lagrange equations) reads, see \cite{GF},
\begin{equation}
    \delta W_{ex}(\mathbf{x}_0,t_0,\mathbf{x},t) = L_{\dot{\mathbf{x}}} \Big|_{t} \cdot \delta \mathbf{x} -
 L_{\dot{\mathbf{x}}_0} \Big|_{t_0} \cdot \delta \mathbf{x}
_0    
    +\left(L  - L_{\dot{\mathbf{x}}}  \cdot \dot{\mathbf{x}} \right)\Big|_{t}\delta t
    -\left(L  - L_{\dot{\mathbf{x}}}  \cdot \dot{\mathbf{x}} \right)\Big|_{t_0} \delta t_0.
\end{equation}
If the initial time and position are fixed, then the variation depends only on $t$ and $x$ and derivatives of the action become
\begin{equation}
   (W_{ex})_{\mathbf{x}}=\mathbf{x}^*,\qquad  
   (W_{ex})_{t} = -H(\mathbf{x},\mathbf{x}^*,t),
\end{equation}
where the momentum is $\mathbf{x}^* = L_{\dot{\mathbf{x}}}$ and the Hamiltonian $H(\mathbf{x},\mathbf{x}^*,t)$ is the Legendre transform of the Lagrangian $L$.

If the Hamiltonian is not explicitly dependent on time, then the second derivative of the Hamilton's principal function with respect to the time vanishes, $(W_{ex})_{tt}=0$, and $W_{ex}$ turns out to be an affine function of $t$. Moreover, also the Lagrangian is then independent of time, the Hamiltonian is conserved and thus equal to a constant, say $\epsilon$. Physically, this constant is identified with the energy of an isolated mechanical system in an inertial reference frame. For the Hamilton's principal function, we then get the Ansatz $W(\mathbf{x},t)=W(\mathbf{x})-\epsilon t$ and the Hamilton-Jacobi equation \eqref{tdepHJ} reduces to its stationary form,
 \begin{equation}\label{hje}
H\left(\mathbf{x},W_\mathbf{x}\right)=\epsilon.
 \end{equation}
Instead of this variational interpretation of the stationary HJ equation, the following Section contains a geometric interpretation.

\textbf{Geometric Hamilton-Jacobi Theory.} Let us now recall the geometric Hamilton-Jacobi theory.  
Consider a Hamiltonian vector field $X_H$ on $T^*M$ and a closed one-form section $\gamma$ (which is locally exact $\gamma=dW$) on $M$.
Together, they define a vector field
\begin{equation}
  X_H\circ \gamma = H_{\mathbf{x}^*}\big|_{\mathbf{x}^* = W_{\mathbf{x}}}\cdot \nabla_{\mathbf{x}}-H_{\mathbf{x}}\big|_{\mathbf{x}^* = W_{\mathbf{x}}}\cdot \nabla_{\mathbf{x}^*},
\end{equation}
which can be projected to $TM$, forming vector field $X_H^{\gamma}$ on $M$,
\begin{equation}\label{gammarelated}
 X_H^{\gamma}:=T\pi\circ X_H\circ \gamma
 = H_{\mathbf{x}^*}\big|_{\mathbf{x}^*=W_{\mathbf{x}}}\cdot \nabla_{\mathbf{x}}.
\end{equation}
The following diagram shows the mappings and vector fields:
\begin{equation}\label{HJ-Clas-pic}
\xymatrix{ T^{*}M\ar[dd]^{\pi_{M}} \ar[rrr]^{X_{H}}& &
&TT^{*}M\ar[dd]_{T\pi_{M}} \\ 
& & &\\
M\ar@/^2pc/[uu]^{\gamma}\ar[rrr]^{X_{H}^{\gamma}}& & & TM\ar@/_2pc/[uu]^{T\gamma}}
\end{equation}
Vector field $X_H^\gamma$ (the bottom arrow) is actually the path from $M$ up to $T^*M$, then to $TT^*M$, and finally down to $TM$.
On the other hand, the vector field $X_H^\gamma$ can be mapped back to $TT^*M$ by the tangent mapping $T\gamma$, 
\begin{equation}
  T\gamma\circ X_H^\gamma = H_{\mathbf{x}^*}\big|_{\mathbf{x}^* = W_{\mathbf{x}}}\cdot \nabla_{\mathbf{x}} +W_{\mathbf{xx}}H_{\mathbf{x}^*}\big |_{\mathbf{x}^* = W_{\mathbf{x}}}  \cdot \nabla_{\mathbf{x}^*},
\end{equation}
which is the path $X_H^\gamma$ in Diagram \eqref{HJ-Clas-pic} mapped up to $TT^*M$. 
Is this vector field equivalent with $X_H\circ\gamma$?
They are equivalent if and only if 
\begin{equation}\label{ext-gr}
  H_{\mathbf{x}}\big|_{\mathbf{x}^* = W_{\mathbf{x}}} + W_{\mathbf{xx}} H_{\mathbf{x}^*}\big|_{\mathbf{x}^*=W_{\mathbf{x}}} =0,
\end{equation}
which means that $d (H\circ \gamma)=0$.  
In summary, we have arrived at the following theorem  \cite{carinena2006geometric}. 
\begin{theorem} \label{HJT}
A closed one-form $\gamma=dW$ on $M$ is a solution of the Hamilton--Jacobi equation \eqref{hje} if the following conditions are satisfied:
\begin{enumerate}
\item The vector fields $X_{H}$ and $X_{H}^{\gamma }$ are $\gamma$-related, that is
\begin{equation}\label{commu}
T\gamma \circ X_H^{\gamma} =X_H\circ\gamma.
\end{equation}
\item Or, equivalently, if the following equation is fulfilled
\begin{equation}\label{2nd}
d\left( H\circ \gamma \right)=0.
\end{equation}
\end{enumerate}
\end{theorem}
\noindent
The second
condition implies that exterior derivative of the Hamiltonian function on the image of $\gamma $ is closed, that is, $%
H\circ \gamma $ is constant. In coordinates, we have that
\begin{equation}
H\left( \mathbf{x},\boldsymbol{\gamma}\left( \mathbf{x}\right) \right) = \epsilon.  \label{HJ-0}
\end{equation}%
The substitution of the local realization $dW=\gamma $ into \eqref{HJ-0} gives the Hamilton-Jacobi equation in form (\ref{hje}). 

In the literature, the geometric Hamilton-Jacobi theorem \ref{HJT} has been extended in various different dynamical formulations. For example, geometric Hamilton-Jacobi theorem has been studied for  higher order systems in \cite{colombo2014geometric}, for field theories in \cite{HJ-field,HJ-co-k-field,HJ-k-field}, for implicit dynamics in \cite{EsLeSa18,EsLeSa20}, for non-holonomic dynamics in \cite{CaGrMaMaMuMiRa,EsJiLeSa19}.  We refer to a recent review \cite{esen2022reviewing} for a more complete picture.  

\textbf{Lift of Solutions.}
Consider a section $\gamma=dW$ satisfying the Hamilton-Jacobi equation and condition \eqref{2nd}. Then this section lifts a solution $(\mathbf{x}(t))$ of the dynamics on $M$ generated by $X^\gamma_H$ to the solution $(\mathbf{x}(t),W_\mathbf{x}(\mathbf{x}(t)))$ of the Hamilton's equation on the cotangent bundle $T^*M$. Such a solution of the Hamiltonian equations is
called horizontal, since it is on the image of a one-form on $M$. Let us now find the evolution equations governing the Hamiltonian flow.

The first set of equations governing the Hamiltonian flow on $TT^*M$ and the dynamics on $M$ are the same. Therefore, it only remains to check the second set of equations in \eqref{HamEqLoc-}. A direct calculation shows that the second set satisfies
 \begin{equation} \label{solution-2-evo-} 
 \frac{d\mathbf{x}^*}{dt} =
 \frac{d}{dt}\left(W_\mathbf{x}\right)=
 ( \dot{\mathbf{x}}\cdot \nabla_\mathbf{x} )W_\mathbf{x}=
 ( H_{\mathbf{x}^*}\big\vert_{\mathbf{x}^*=W_\mathbf{x}}\cdot \nabla_\mathbf{x} )W_\mathbf{x}=-H_{\mathbf{x}}\big\vert_{\mathbf{x}^*=W_\mathbf{x}},
 \end{equation}
where we have employed \eqref{ext-gr}. Note that \eqref{solution-2-evo-} is precisely the second equation in the Hamilton's equations \eqref{HamEqLoc-}. 
    
\textbf{A Simple Illustrative Example.}    Let us illustrate the cotangent lift and Hamilton-Jacobi equation on a concrete and simple example.  
Consider a particle with position $\xx=\qq$ and momentum $\xx^*=\pp$. The total energy of the particle in a potential $V$ is $H=\Vert\mathbf{p}\Vert^2/2m + V(\mathbf{q})$. Assume that $W(\mathbf{q}) = -\zeta V(\qq)$ is determining the exact one-form $\gamma(\qq) = (\qq,-\zeta V_{\qq})$. The vector field $X_H^\gamma$ represents a dissipative evolution of the position $\qq$ towards the equilibrium (minimum of potential $V(\qq)$),
 \begin{equation}\label{exp-proj-v-f}
 X_H^\gamma=  -\frac{1}{m}\zeta V_{\qq} \cdot \nabla_{\qq},\qquad 
 \frac{d \qq}{dt} = -\frac{1}{m}\zeta V_{\qq}.
\end{equation} 
The tangent map of this vector field is 
\begin{equation}
T\gamma \circ X_H^\gamma  =  -\frac{1}{m}\zeta V_{\qq}\cdot \nabla_{\qq} + \frac{1}{m}\zeta^2 V_{\qq\qq} V_{\qq} \cdot \nabla_{\pp}.
\end{equation}  
This dynamics is equivalent with dynamics \begin{equation}
X_H\circ\gamma =  (-\frac{1}{m}\zeta V_\qq, -V_\qq)
\end{equation}  
 if and only if 
\begin{equation}\label{eq.HJ.compat.1}
  0 = d(H\circ\gamma) = -\frac{1}{m}\zeta^2 V_{\qq\qq} V_{\qq}  + V_\qq, 
\end{equation}
which is satisfied for $V(\qq) = (m/2\zeta^2)\Vert\qq\Vert^2$. In other words, for such potential, the reduced evolution given by vector field $X_H^\gamma$ is equivalent to the restriction of the Hamiltonian vector field to section $\gamma$, that is $X_H\circ\gamma$, by applying the tangent map $T\gamma$ to $X_H^\gamma$.

\subsection{Morse Families and Complete Solutions}
\label{sec-Morse}

The lifts presented in the previous subsection provide solutions to the Hamilton's equation that are horizontal (first computed on $n$-dimensional manifold $M$ and then lifted to the $2n$-dimensional cotangent bundle $T^*M$). But such solutions do not contain trajectories that are inherently $2n$-dimensional (complete solutions) and, in order to arrive at complete solutions, we need $2n$ independent variables. The concept of Morse families provides a geometric way towards the complete solutions.

Before the Morse families, let us recall non-horizontal Lagrangian submanifolds, which play a key role in the definition of Morse families.
A submanifold of $T^{\ast}M$ is called Lagrangian if the symplectic form vanishes on it and if it is maximal with this property. 
For instance, let $F$ be a real valued function on a manifold $M$. Then the image of its exterior derivative
\begin{equation}
dF:M\longrightarrow T^{\ast }M,\qquad 
\mathbf{x}\mapsto \left( \mathbf{x,}F_{\mathbf{x}}\right)   
\end{equation}%
determines a Lagrangian submanifold of the symplectic manifold $T^{\ast
}M$. The image of the function can be depicted as follows:
 \begin{equation} \label{LagLag}
  \xymatrix{ T^*M \ar[dd]^{\pi_M}  \\ \\ 
M  \ar@/^2pc/[uu]^{dF}
}
\end{equation}
A Lagrangian submanifold of $T^{\ast }M$ is called non-horizontal if it cannot be written as image of a
section of the fibration $T^{\ast }M \mapsto M$. For instance, consider a circle (Lagrangian submanifold) in a plane (cotangent bundle over a line). There is no smooth function that maps points of the line to the points of the circle, which means that the circle is non-horizontal.

\textbf{Morse Families.}
A generating function $F$ defined on $M $ can only have $n$ variables. To increase the number of independent variables, we use the concept of Morse families. Roughly speaking, one can say that a Morse family is formed by functions on $M $ depending on some additional variables with the property that the Lagrangian submanifold generated by the Morse family is non-horizontal. 

Consider a real valued function $E$
defined on the total space of a smooth bundle $(Y,\tau ,M)$ over the state space $M$. If the 
coordinates on $Y$ are $\left( \mathbf{x,y}\right)$, then $E$ is called a Morse
family when the rank of the matrix 
\begin{equation*}
rank\left( E_{\mathbf{xy}}\text{ \ \ }E_{\mathbf{yy}}\right)
\end{equation*}%
is maximal. 
The submanifold  
\begin{equation} \label{Lag-Morse}
L_{E}=\left\{ \left( \mathbf{x},\mathbf{x}^*\right) \in T^{\ast }M :~\mathbf{x}^*=E_{\mathbf{x}}(\mathbf{x,y}),~ E_{\mathbf{y}}(\mathbf{x,y})=0\right\}
\end{equation}%
is then a Lagrangian submanifold of $T^{\ast }M$, and the following diagram shows the geometric structure of Morse families:
 \begin{equation} \label{Morse-pre-}
  \xymatrix{
\mathbb{R}& Y \ar[dd]^{\tau}\ar[l]_{E}& T^*M \ar[dd]^{\pi_M} \ar@(ul,ur)^{L_{E}}  & \\ \\ &
M  \ar@{=}[r]& M &
}
\end{equation}

One may generalize this picture by considering a fibration over a submanifold of the base manifold $M$,
 \begin{equation} \label{Morse-pre}
  \xymatrix{
\mathbb{R}& Z \ar[dd]^{\tau}\ar[l]_{E}& T^*M \ar[dd]^{\pi_M} \ar@(ul,ur)^{L_{E}}  & \\ \\ &
N  \ar@{^{(}->}[r]& M &
}
\end{equation}
where $N$ is a submanifold embedded into $M$ and $(Z,\tau,M)$ stands for a fiber-bundle structure on $N$. Function $E$ is then a function on the total space $Z$. By a slight abuse of notation, we can consider $\mathbf{a}$ as coordinates on $N$ while taking $\mathbf{x}=(\mathbf{a},\mathbf{b})$ as local coordinates on $M$. When coordinates on the total space $Z$ are $(\mathbf{a},\mathbf{u})$, the Morse family condition becomes the maximality of $ 
rank\left( E_{\mathbf{au}}\text{ \ \ }E_{\mathbf{uu}}\right)$. In this realization, the Lagrangian submanifold generated by $E$ is 
 \begin{equation} \label{Morse-pre-gen}
L_{E}=\left\{ \left( \mathbf{a},\mathbf{b},\mathbf{a}^*,\mathbf{b}^*\right) \in T^{\ast }M :~ \mathbf{a}^*=E_{\mathbf{a}}(\mathbf{a,u}),~ \mathbf{b}^*=0,~ E_{\mathbf{u}}(\mathbf{a,u})=0\right\}.
\end{equation}
Finally, all non-horizontal Lagrangian submanifolds admit such a local characterization, which is called Maslov-Hörmander theorem (or generalized Poincar\'{e} Lemma), see \cite{Benenti-book,Liber87,Tulczyjew-slow}. 

\textbf{Complete Solutions.}
To describe also complete solutions of Hamilton's equation, we have to merge Morse families and the Hamilton-Jacobi Theorem \ref{HJT}, replacing the role of  $\gamma$ with a Morse family $E$, as in the following diagram: 
\begin{equation}\label{HJ-Clas-pic-Morse}
\xymatrix{ 
\mathbb{R}& Y \ar[dd]^{\tau}\ar[l]_{E}& T^{*}M\ar@(ul,ur)^{L_{E}}\ar[dd]^{\pi_{M}} \ar[rrr]^{X_{H}}& &
&TT^{*}M\ar[dd]_{T\pi_{M}} \\ & & &\\ &
M  \ar@{=}[r]& 
M \ar[rrr]^{X_{H}^{L_{E}}}& & & TM}
\end{equation}
The Lagrangian submanifold $L_{E}$ is generated by a Morse family $E$ defined on the bundle $(Y,\tau,M)$. 

First we restrict the Hamiltonian vector field $X_H$ to the Lagrangian submanifold $L_E$ and then we project the restriction to the tangent bundle $TM$. The restriction of 
$X_H$ is 
\begin{equation}  \label{HamEqLoc--}
X_{H}\big\vert_{L_E} (\mathbf{x},\mathbf{y})= H_{\mathbf{x}^*}(\mathbf{x},E_\mathbf{x}) \cdot \nabla_\mathbf{x} - H_{\mathbf{x}}(\mathbf{x},E_\mathbf{x}) \cdot \nabla_{\mathbf{x}^*},\qquad E_\mathbf{y}=0,
\end{equation}  
and the projection $T\pi_{M}\circ X_{H}\big\vert_{L_E}$ is a submanifold of $TM$. Then, the projected dynamics becomes
\begin{equation}  \label{HamEqLoc-impl}
\frac{d\mathbf{x}}{dt}=H_{\mathbf{x}^*}(\mathbf{x},E_\mathbf{x}(\mathbf{x,y})), \qquad E_\mathbf{y}(\mathbf{x,y})=0.
\end{equation}  
Moreover, if the dimension of the fibers in the fibration $\tau$ is $n$, then the solutions of implicitly defined dynamics \eqref{HamEqLoc-impl} involve $2n$ variables. By lifting 
these solutions to the cotangent bundle $T^*M$ by means of the Morse family, we arrive at the complete solutions of Hamilton's equation. 

Note also that the Morse family determines a symplectic diffeomorphism 
 \begin{equation}
 T^*M\longrightarrow T^*M,\qquad (\mathbf{x},E_\mathbf{x}(\mathbf{x,y}))\mapsto (\mathbf{y},-E_\mathbf{y}(\mathbf{x,y})). 
 \end{equation}  


\subsection{Holonomic Lift and Vertical Representative}\label{Rgeq-geometry}
In the preceding text we have recalled geometric lifts of vector fields, and in the present Section, we summarize how the lifted fields can be geometrically split into two parts (holonomic and vertical). Such splitting allows for instance to reduce a detailed dynamics to a less detailed description. But first we have to recall the concepts of jet bundles and jet decomposition.

\textbf{Jet Bundle.} Consider a fiber bundle $(P,\pi ,M)$ with base coordinates $\mathbf{x}=(x^i)$ on $M$ and induced coordinates $(\mathbf{x},\mathbf{u})=(x^i,u^\lambda)$ on the total manifold $P$. Two local sections are called 1-equivalent at $\mathbf{x}$ if 
\begin{equation}
		\phi(\mathbf{x})=\psi(\mathbf{x}),\qquad \phi_\mathbf{x}(\mathbf{x})=\psi_\mathbf{x}(\mathbf{x}),
	\end{equation} 
and the set of 1-equivalent sections forms an equivalence class called 1-jet of $\phi$, denoted by $j_x^1 \phi$ \cite{Saunders-book}. 

Moreover, the set of 1-jets $j_x^1 \phi $ forms a smooth manifold called first jet manifold, denoted by $J^1\pi$. The induced coordinates on $J^1\pi$ are $(\mathbf{x},\mathbf{u},\mathbb{U})=(x^i,u^\lambda,u^\lambda_i)$. 
The first jet manifold $J^1\pi$ has two projection maps, $\pi_1$ and $\pi_{1,0}$, 
\begin{align*}
	  \pi_1&: J^1\pi \to M   :   j_x^1\phi \to \mathbf{x}  \\
	 \pi_{1,0}&:J^1\pi \to P  :   j_x^1\phi \to \phi(\mathbf{x} ),
\end{align*}
as in the following diagram:
\begin{equation}
\begin{tikzcd}
	J^1\pi \arrow{rr}{\pi_{1,0}} \arrow[swap]{ddr}{\pi_1} & & P \arrow{ddl}{\pi} \\ \\
	& M
\end{tikzcd}
\end{equation} 

Finally, $J^1\pi $ is the smooth bundle with projection $\pi_1$, and $(J^1\pi, \pi_1, M)$ is called the first jet bundle of $\pi$. For a real valued function $\phi$ on the base manifold, a section of the fibration $J^1\pi \to M$ is locally in the form of
\begin{equation}\label{jet-sect}
J\phi:M\longrightarrow J^1\pi,\qquad  \mathbf{x} \mapsto  (\mathbf{x},\phi_\mathbf{x}(\mathbf{x}),\phi_\mathbf{xx}(\mathbf{x})),
\end{equation}
where $\phi_\mathbf{xx}$ is the Hessian matrix.

\textbf{Holonomic Lift.} 
Consider a fiber bundle $(P,\pi ,M)$ with base coordinates $\mathbf{x}=(x^i)$ on $M$ and coordinates $(\mathbf{x},\mathbf{u})=(x^i,u^\lambda)$ on the total space.
Let $X$ be a vector field on $M$ and $\sigma$ a section of the fibration $\pi$. Then, the Lie derivative (directional derivative) of a smooth function $F$ defined on the total space $P$ with respect to the vector field $X$ can be computed by means of $\sigma$ as 
$\mathcal{L}_X(F\circ \sigma)$. 
This leads to the definition of the holonomic lift,
\begin{equation}
X^{hol}:J^1\pi\longrightarrow TP
\end{equation}
of the vector field $X$ to the total space $P$ of the fibration, 
\begin{equation} \label{hol-lift}
X^{hol}(F)\circ \phi:=X(F\circ \phi),
\end{equation}
see \cite{esen2019lifts,esgum11,olver86,Saunders-book}.
Note that $X^{hol}(F)$ stands for the directional derivative of the function $F$ in the direction of the vector field $X^{hol}$ whereas $X(F\circ \phi)$ is the directional derivative of $F\circ \phi$ in the direction of $X$. 

In terms of the local coordinates, if $X=\mathbf{X}\cdot \nabla_{\mathbf{x}}$, the holonomic lift (\ref{hol-lift}) becomes
\begin{equation}\label{hol-lift-defn}
X^{hol}(\mathbf{x},\mathbf{u},\mathbb{U})=\mathbf{X}\cdot \nabla_{\mathbf{x}} + 
\mathbb{U}\mathbf{X}\cdot \nabla_{\mathbf{u}}=X^{i}\frac{\partial }{\partial x^{i}}+X^{i}u_{i  }^{\lambda }\frac{\partial 
}{\partial u^{\lambda }}.
\end{equation}
Note, however, that $X^{hol}$ is not a classical vector field on $P$, since its coefficients depend on the first order jet bundle term $\mathbb{U}=[u_{i}^{\lambda }]$. 
Instead, it is called a generalized vector field. 

Finally, the following diagram demonstrates the geometric construction of holonomic lifts, 
\begin{equation}\label{Ham-GENERIC}
\xymatrix{ J^1 \pi  \ar[dd]^{\pi_{1}} \ar[rrr]^{X ^{hol}}& &
&TP\ar[dd]^{T\pi_{1}} \\ & & &\\
M\ar[rrr]^{X}& & & TM}
\end{equation}
In order to justify the term holonomic, note that the values of $X^{hol}$ at the contact one-forms 
\begin{equation}
\boldsymbol{\vartheta}=d\mathbf{u} - \mathbb{U} d\mathbf{x},\qquad \vartheta^\lambda=du^\lambda- u_{i  }^{\lambda }dx^i,
\end{equation}
 vanish identically. 
 
\textbf{Vertical Representative.} Consider once more a fiber bundle $(P,\pi,M)$ and also a projectable vector field $Y$ on the total space $P$, which we can be projected to the base manifold $M$ by the fiber bundle projection $\pi$ to a vector field $\pi_*Y$. Then, the holonomic lift of $\pi_*Y$ gives a generalized vector field $(\pi_*Y)^{hol}$ with values in $TP$, called the holonomic part of the vector field $Y$ and denoted by $\mathfrak{H}Y:=(\pi_*Y)^{hol}$. Finally, the difference between the vector field $Y$ and its holonomic part $\mathfrak{H}Y$ is the vertical representative of $Y$, denoted by $VY$. 

In coordinates, for a projectable vector field 
\begin{equation}
Y(\mathbf{x},\mathbf{u})=\mathbf{Y}(\mathbf{x})\cdot \nabla_\mathbf{x} + \mathbf{Z}(\mathbf{x},\mathbf{u})\cdot \nabla_\mathbf{u},
\end{equation}
the holonomic part and the vertical representative are 
\begin{equation}
\mathfrak{H}Y(\mathbf{x},\mathbf{u},\mathbb{U})=\mathbf{Y}\cdot \nabla_{\mathbf{x}} + 
\mathbb{U}\mathbf{Y}\cdot \nabla_{\mathbf{u}},\qquad \mathfrak{V}Y(\mathbf{x},\mathbf{u},\mathbb{U})=(\mathbf{Z}-
\mathbb{U}\mathbf{Y})\cdot \nabla_{\mathbf{u}}.
\end{equation}

Although both $\mathfrak{H}Y$ and $\mathfrak{V}Y$ are generalized vector fields, their sum is a classical vector field, since $Y$ does not depend on the jet coordinate $\mathbb{U}$. Now we are in position to apply these geometric concepts in Hamiltonian dynamics and non-equilibrium thermodynamics.

\subsection{Jet Decomposition of Hamiltonian Dynamics} 
Jet decomposition of Hamiltonian dynamics is a geometric way splitting Hamiltonian vector fields into two components, the holonomic part and the vertical representative. The latter will play the role of fast dynamics on top of the slower holonomic part.

Let us consider a cotangent bundle $\pi_M:T^*M\mapsto M$. If $M$ is $n$-dimensional, then the first jet bundle $J^1\pi_M$ has dimension $2n+n^2$, and in the local coordinates it becomes
\begin{equation}
(\mathbf{x},\mathbf{u},\mathbb{U})=(\mathbf{x},\mathbf{x}^*,\mathbf{x}^*_\mathbf{x}),
\end{equation}
where the jet fiber coordinates are $\mathbb{U}=\mathbf{x}^*_\mathbf{x}=[ \partial x^*_i / \partial x^j ]$. Now, we employ the holonomic lift to a vector field $X=\mathbf{X}\cdot \nabla_\mathbf{x}$ on $M$, using the cotangent-bundle projection. This copies the flow on $M$ to a flow on the cotangent bundle,
\begin{equation}
 X ^{hol}(\mathbf{x},\mathbf{x}^*,\mathbf{x}^*_{\mathbf{x}})=\mathbf{X}\cdot \nabla_\mathbf{x} +\mathbf{x}^*_\mathbf{x}
\mathbf{X} \cdot \nabla_{\mathbf{x}^*},
 \end{equation}
where $\mathbf{x}^*_\mathbf{x} \mathbf{X}$ stands for matrix multiplication of $\mathbf{x}^*_\mathbf{x}$ and the vector $\mathbf{X}$. The dynamics generated by this holonomic lift is
\begin{equation} \label{idento-2}
 \frac{d \mathbf{x} }{dt} = \mathbf{X} ,\qquad 
 \frac{d \mathbf{x}^*}{dt} =\mathbf{x}^*_\mathbf{x}
\mathbf{X}.
\end{equation}

Consider now a Hamiltonian vector field $X_H$ on the cotangent bundle $T^*M$ and the projected vector field $X_H^\gamma$ on the base manifold $M$, defined in \eqref{gammarelated}. First, we shall calculate the holonomic lift of $X_H^\gamma$ to the cotangent bundle and, subsequently, we will the decompose the Hamiltonian vector field $X_H$ into the sum of its holonomic part and vertical representative. Assuming the local identification $\gamma=dW$, the local realization of the projected field $X_H^\gamma$ can be found in Equation \eqref{gammarelated}. 
The holonomic lift of the projected dynamics is then
  \begin{equation}
 \mathfrak{H}X_H(\mathbf{x},\mathbf{x}^*,\mathbf{x}^*_{\mathbf{x}}):= (X_H^\gamma)^{hol}(\mathbf{x},\mathbf{x}^*,\mathbf{x}^*_{\mathbf{x}})=
H_{\mathbf{x}^*}\big\vert_{\mathbf{x}^*=W_{\mathbf{x}}}\cdot \nabla_{\mathbf{x}} +  \mathbf{x}^*_\mathbf{x}
H_{\mathbf{x}^*}\big\vert_{\mathbf{x}^*=W_{\mathbf{x}}}\cdot \nabla_{\mathbf{x}^*}.
   \end{equation}
Note that the holonomic lift $(X_H^\gamma)^{hol}$ allows to define the holonomic part of the Hamiltonian vector field $X_H\big\vert_{im \gamma}$ restricted to the image of $\gamma$ as $\mathfrak{H}X_H$. The complementary vector field is the vertical representative,
\begin{equation}
   \mathfrak{V}X_H(\mathbf{x},\mathbf{x}^*,\mathbf{x}^*_{\mathbf{x}})=\big(-H_{\mathbf{x}}\big \vert_{\mathbf{x}^*=W_{\mathbf{x}}} 
    -  \mathbf{x}^*_\mathbf{x}
H_{\mathbf{x}^*}\big\vert_{\mathbf{x}^*=W_{\mathbf{x}}}\big)\cdot \nabla_{\mathbf{x}^*},
  \end{equation}  
which means that we have the following holonomic and vertical decomposition:
\begin{equation}
X_H= \mathfrak{H}X_H+\mathfrak{V}X_H.
  \end{equation} 
  
A direct computation gives that the holonomic part $\mathfrak{H}X_H$ is $\gamma$-\textit{related} with the projected field $X_H^{\gamma}$,
\begin{equation}\label{hol-com}
T\gamma \circ X_H^{\gamma}=\mathfrak{H}X_H \circ J^1\gamma,
\end{equation}
where $J^1\gamma$ is the first jet prolongation of the section $\gamma$. The following diagram summarizes this commutation property, 
\begin{equation}\label{Ham-GENERIC-2}
\xymatrix{J^1\pi_M \ar[rrr]^{\mathfrak{H}X_H}& &
&TT^*M  \\ & & &\\
M \ar[uu]^{J^1\gamma} \ar[rrr]^{X_H^{\gamma}}& & & TM \ar[uu]_{T\gamma}}
\end{equation}
where $T\gamma$ is the tangent mapping of $\gamma$.

Notice that, according to \eqref{hol-com}, we can replace the left hand side of the equality in \eqref{commu} with $\mathfrak{H}X_H \circ J^1\gamma$. Therefore, we may replace the Hamilton-Jacobi condition \eqref{commu} with an analogical condition for the holonomic lift of the Hamiltonian vector field, which is summarized in the following restatement of geometric Hamilton-Jacobi theorem.

\begin{proposition} \label{HJT-vert}
For a closed one-form $\gamma$ (that is locally $=dW$) on $M$, the following conditions are equivalent:
\begin{enumerate}
\item The identity is fulfilled
\begin{equation}\label{commu-1}
\mathfrak{V}X_H \circ J^1\gamma=0.
\end{equation}
\item The identity is fulfilled
 \begin{equation}
\mathfrak{H}X_H\circ J^1\gamma =X_H\circ \gamma  . 
\end{equation}
\item The identity is fulfilled
\begin{equation}\label{2nd-1}
d\left( H\circ \gamma \right)=0.
\end{equation}
\end{enumerate}
\end{proposition}

The decomposition given in this theorem has interesting implications. First, under condition \eqref{2nd-1}, the holonomic lift is precisely copying the flow from the base level $M$ to the cotangent bundle $T^*M$, and an integral curve of $X_H^{\gamma}$ is lifted immediately to an integral curve of $X_H$. These two flows are in one-to-one relation because of the $\gamma$-relatedness condition \eqref{commu}. Second, if the flows are not copies of each other, the vertical representative is responsible for such incompatibility. Let us demonstrate such incompatibility on a physically motivated example, where a detailed evolution is reduced to a less detailed.
 
\textbf{Revisiting the Simple Illustration.} Consider again $\xx=\qq$ and $\xx^*=\pp$ as position and momentum of a particle with the total energy  $H=\Vert\mathbf{p}\Vert^2/2m + V(\mathbf{q})$. Choosing $W=-\zeta V(\qq)$, the vertical representative becomes 
\begin{equation}
\mathfrak{V}X_H\circ J^1\gamma =- \frac{1}{m} \left(\zeta V_\qq  + \zeta^2 V_{\qq\qq}  V_\qq \right)\cdot \nabla_\pp
,
\end{equation}
which is the same expression as we have obtained in the compatibility condition within the Hamilton--Jacobi theory \eqref{eq.HJ.compat.1}. What if the compatibility condition is not satisfied? Then the lift of the reduced dissipative dynamics is incompatible with the composition of the Hamiltonian vector field and the section $\gamma$. In other words, the more detailed Hamiltonian vector field contains more information than the tangent lift of the reduced dynamics, which can be expected from the physical point of view. 

\textbf{Geometric reduction.} One can also start from the Hamiltonian vector field $X_H$. Section $\gamma = dW$ then maps positions (even with respect to the time-reversal transformation) to the momenta (odd) \cite{pre15}. If, instead of the compatibility condition, we set the vertical representative of the Hamiltonian vector field to zero, 
\begin{equation}
H_\qq\big|_{\pp = W_\qq} = W_{\qq\qq} \cdot  H_{\pp}\big|_{\pp=W_\qq},
\end{equation}
 then the natural projection of the Hamiltonian vector field becomes 
\begin{equation}
\dot{\qq}=T\pi_M \circ X_H\circ\gamma = W^{-1\, \qq\qq}H_{\qq} = -\zeta^{-1}   V_{\qq},
\end{equation}
 where $W^{-1\, \qq\qq}$ is the inverse Hessian of $W$. 
In other words, a section generated by $W=-\zeta V(\qq)$ and vanishing vertical representative reduce the reversible Hamiltonian vector field (dynamics for $\qq$ and $\pp$) to irreversible dynamics for $\qq$, approaching the minimum of potential $V(\qq)$. 

\section{On Geometry of Irreversible Motion}\label{3}
In this Section, we reformulate irreversible gradient dynamics as Hamiltonian dynamics, using the Hamilton-Jacobi theory discussed above. This provides a geometric unification for the two parts of the GENERIC framework, the Hamiltonian part and the gradient part.

\subsection{Dissipation Potential and Gradient Dynamics} \label{Grad-Sec}
First we recall gradient dynamics, generated by a dissipation potential \cite{gyarmati,Otto}.
Consider a symplectic bundle $(T^*M,\Omega_M)$ equipped with the Darboux' coordinates $(\mathbf{x},\mathbf{x}^*)$. 
The dissipation potential $\Xi=\Xi(\mathbf{x},\mathbf{x}^*)$ is a real-valued function on $T^{*}M$ satisfying the following local properties \cite{go,og}:
\begin{itemize}
\item $\Xi(\mathbf{x},0)=0$ for all $\mathbf{x}$.
\item $\Xi$ reaches its minimum at $\mathbf{x}^*=0$. 
\item $\Xi$ is a convex function of $\mathbf{x}^*$ in a neighborhood of $\mathbf{x}^*=0$ for all $\mathbf{x}$. 
\end{itemize}
Our aim is to see the dissipation potential $\Xi$ as a Hamiltonian function on the cotangent bundle $T^*M$, which allows to compute the associated
Hamiltonian vector field $X_{\Xi}$ according to the local formula \eqref{HamEqLoc-}.

\textbf{Gradient Flow.} Consider a real-valued function  $S$ on the base manifold $M$ called entropy. The exterior derivative 
 $dS$ is a section of the cotangent bundle,
\begin{equation}
dS:M\longrightarrow T^{*}M ,\qquad (\mathbf{x})\mapsto (\mathbf{x},\mathbf{x}^*=S_\mathbf{x}).
\end{equation}
Then, considering the commutative diagram 
\begin{equation}\label{Xg}
  \xymatrix{ 
  T^{*}M
\ar[dd]^{\pi_{M}} \ar[rrr]^{X_{\Xi}}&   & &TT^{*}M\ar[dd]^{T\pi_{M}}\\
  &  & &\\
 M\ar@/^2pc/[uu]^{dS}\ar[rrr]^{X_{\Xi}^{dS}}&  & & TM }
\end{equation}
we project down the Hamiltonian flow $X_{\Xi}$ to a flow on the base manifold $M$ as 
\begin{equation}  \label{gammarelated-}
X_{\Xi}^{dS}:M\longrightarrow TM,\qquad X_{\Xi}^{dS}=T\pi_M \circ
X_{\Xi}\circ dS.
\end{equation}
In a local chart, the dynamics governed by the projected vector field $X_{\Xi}^{dS}$ becomes
\begin{equation}  \label{grad}
X_{\Xi}^{dS}=\Xi_{\mathbf{x}^*}
\big\vert_{\mathbf{x}^*=S_\mathbf{x}}\cdot \nabla _\mathbf{x},\qquad 
\frac{d \mathbf{x}}{dt}= \Xi_{\mathbf{x}^*}
\big\vert_{\mathbf{x}^*=S_\mathbf{x}}.
\end{equation}
In other words, gradient flow can be seen as a part of a Hamiltonian flow on the cotangent bundle.

Potentials $\mathbb{C}= \mathbb{C}(x)$ for which
\begin{equation} 
X_{\Xi}^{dS} (\mathbb{C})= \left.\frac{\partial \Xi}{\partial x^*_i}
\right\vert_{x^*=S_x} \frac{\partial \mathbb{C}} {\partial x^i}=0, \qquad X_{\Xi}^{d\mathbb{C}} = 0 
\end{equation}
are called dissipation Casimirs. From the physics point of view, energy will be a dissipation Casimir so that the first law of thermodynamics (conservation of energy) is satisfied.

\textbf{Gradient Flow versus Hamiltonian Flow.}
Since $X_\Xi$ is a vector field on a higher-dimensional manifold than $X_{\Xi}^{dS}$, it can contain more information. However, the two vector fields can also be $\gamma$-related, and then they are compatible with each other. This is summarized in the following variant of the geometric Hamilton-Jacobi Theorem \ref{HJT}.

\begin{proposition}
The Hamiltonian dynamics $X_{\Xi}$ in
\eqref{HamEqLoc} and the gradient dynamics $
X_{\Xi}^{dS} $ in \eqref{grad} are compatible in the sense that
\begin{equation}
    TdS(X_{\Xi}^{dS})=X_{\Xi} \circ dS.\label{eq.compat}
\end{equation}
if and only if the entropy is a solution of the Hamilton-Jacobi equation
\begin{equation}
\Xi
\big\vert_{\mathbf{x}^*=S_\mathbf{x}}=\epsilon
\end{equation}
for a constant $\epsilon$. 
\end{proposition}
If either of these conditions is satisfied, the exterior derivative of the entropy lifts the integral curves of the gradient vector field $X_{\Xi}^{dS}$ to integral curves of the Hamiltonian vector field $X_{\Xi}$, and the Hamiltonian lift of the gradient dynamics is then equivalent with the gradient dynamics itself. This feature can be used for instance when integrating the gradient dynamics, since one can then use integrators for Hamiltonian mechanics like symplectic integrators \cite{leimkuehler}.
On the other hand, if the Hamilton-Jacobi condition is not satisfied, the projection still remains valid, but the lift of the vector field \eqref{grad} on $M$ does not constitute a Hamiltonian vector field on $T^*M$.

\textbf{Principle of Least Dissipation.} In terms of the variational principles, the Hamilton-Jacobi equation is a result of criticality of the action integral 
\begin{equation}
    \int_{t_0}^{t} \Xi^*(\mathbf{x},\dot{\mathbf{x}}) dt
\end{equation}
with fixed initial time and point while keeping the final time and point undetermined. Here, $\Xi^*$  is the Lagrangian function obtained by the Legendre transformation of the Hamiltonian function (dissipation potential) $\Xi$ with respect to the fiber variable $\mathbf{x}^*$.
Therefore, compatibility condition \eqref{eq.compat} can be seen also as the principle of least dissipation (assuming $\Xi^*$ convex) \cite{Onsager}.

\textbf{Example: Quadratic Potential.}
Consider for instance a quadratic dissipation potential 
\begin{equation}
\Xi(\mathbf{x},\mathbf{x}^*)=\frac{1}{2}\mathbf{x}^*\cdot\Lambda(\xx)  \mathbf{x}^*
\end{equation}
where $\Lambda(\xx)=[\Lambda^{ij}(\xx)]$ is a symmetric positive (semi-)definite operator. This is the setting of GENERIC with dissipative brackets \cite{hco}, also called metriplectic \cite{mor}. 
The dynamics on the base manifold generated by an entropy $S$ is
\begin{equation}  
\frac{d \mathbf{x}}{dt}= \Lambda S_{\mathbf{x}}.
\end{equation}
Finally, the Hamilton-Jacobi condition becomes
\begin{equation}
\epsilon = \frac{1}{2}S_\xx\cdot\Lambda S_\xx,
\end{equation}
which means that the gradient and Hamiltonian flows are compatible for $\Lambda\propto (S_\xx)^{-2}$. When we take $\xx$ as the field of local energy density, then the Hamilton-Jacobi condition leads to the Fourier law of heat conduction, where heat flux negative temperature gradient multiplied by a constant heat conductivity.
Moreover, the inverse Legendre transformation $\dot{\mathbf{x}}=\Xi_{\mathbf{x}^*}$ is a local diffeomorphism and one has $\Xi^*(\mathbf{x},\dot{\mathbf{x}})=(1/2)\dot{\mathbf{x}} \cdot \Lambda^{-1} \dot{\mathbf{x}}$. 

\textbf{Example: Chemical Reaction.} Let $M$ be a two dimensional manifold with local coordinates $(x,y)$ and consider a simple chemical reaction $x\leftrightarrow y$ (for instance transition between two isomers with concentrations $x$ and $y$). The dissipation potential driving chemical reactions is \cite{grchemkin,mielke-potential}
\begin{equation} 
\Xi = k \sqrt{xy}\cosh\left(\frac{y^*-x^*}{2}\right),
\end{equation} 
which is a function on the cotangent bundle $T^*M$ with coordinates $(x,y,x^*,y^*)$. The Hamiltonian dynamics $X_{\Xi}$ generated by $\Xi$ is 
\begin{equation} \label{chem-kin-ham}
\begin{split}
\frac{dx}{dt}&=-\frac{k\sqrt{xy}}{2}\sinh\left(\frac{y^*-x^*}{2}\right),\\ \frac{dy}{dt}&=\frac{k\sqrt{xy}}{2}\sinh\left(\frac{y^*-x^*}{2}\right), \\
\frac{dx^*}{dt}&= \frac{k\sqrt{y}}{2\sqrt{x}} \cosh\left(\frac{y^*-x^*}{2}\right),\\
\frac{dy^*}{dt}&= \frac{k\sqrt{x}}{2\sqrt{y}} \cosh\left(\frac{y^*-x^*}{2}\right).
\end{split}
\end{equation} 
With the standard entropy for ideal mixtures \cite{pkg}
\begin{equation} 
S(x,y) = x (\ln x -1) + y(\ln y -1),
\end{equation} 
defined on the manifold $M$, the projected dynamics $X_{\Xi}^{dS}$ in \eqref{grad} becomes
\begin{equation}\label{chemkin-lin}
 \frac{dx}{dt}=  \frac{k}{4} (y-x), \qquad \frac{dy}{dt}=  \frac{k}{4} (x-y)
\end{equation}
which represents the law of mass action \cite{Guldbergwaage}. 

Is this dissipation potential constant when composed with the section $\gamma=dS$? By plugging $dS$ into $\Xi$, we obtain that $\Xi\circ dS = k (x+y)$, which is indeed a constant (constant mass of the system). Chemical kinetics thus satisfies the Hamilton-Jacobi compatibility condition, which makes it possible to study dynamics of reacting systems by means of Hamiltonian mechanics and related techniques (for instance symplectic integrators). Conversely, dynamics \eqref{chemkin-lin} is linear and can be solved easily, which means that due to the Hamilton-Jacobi theory, we can simplify the problem of solving Hamiltonian system \eqref{chem-kin-ham}.

\subsection{GENERIC as a Projection of Hamiltonian Flow}\label{sectiongeneric}
Now we are finally prepared to formulate a geometric unification of the Hamiltonian and gradient parts of the GENERIC framework.
In Section \ref{Grad-Sec}, irreversible flow is formulated as a projection of a
Hamiltonian flow on the cotangent bundle while Section \ref{Sec-Poisson} contains a
geometric lift of Hamiltonian mechanics to the cotangent
bundle. We merge these two approaches to arrive at a geometric formulation of GENERIC. 

Let $M$ be a Poisson manifold equipped with a Poisson bivector $\mathbb{L}$, and consider a dissipation potential $\Xi$ on the cotangent bundle $T^*M$. Hamiltonian mechanics on $M$ is generated by the Poisson bivector and an energy function $E$ while the irreversible motion is generated by a dissipation potential $\Xi$ and entropy $S$. We assume that the entropy is a concave function of $\mathbf{x}$.
Both geometrical structures are required to be complementary and degenerate in the sense that
\begin{equation}\label{degg}
\mathbb{L}  S_{\mathbf{x}} =0, \qquad  
X_{\Xi}^{dS} (E)=0, \qquad X_{\Xi}^{dE} = 0.
\end{equation}
The degeneracy  (\ref{degg}) is thus the requirement
that the entropy $S$ be a Casimir of the Poisson bracket whereas the energy $E$ be a dissipation Casimir of the gradient flow. The thermodynamic potential is a formulated as a linear combination of the entropy and the energy,
\begin{equation}\label{Phi}
\Phi(\mathbf{x},e^*)=-S(\mathbf{x})+e^*E(\mathbf{x}),
\end{equation}
where $e^*$ is a Lagrange multiplier in the maximization of the entropy $S $ subjected to the constraint $E $. In terms of the terminology and geometry of Subsection \ref{sec-Morse}, we call $\Phi$ a Morse family defined on the total space of the line bundle $M\times \mathbb{R}$ over $M$. The coordinates on $M\times \mathbb{R}$ are $(\mathbf{x},e^*)$. According to the formulation \eqref{Lag-Morse}, the Lagrangian submanifold of $T^*M$ generated by $\Phi$ is 
 \begin{equation}
 L_\Phi=\{(\mathbf{x},\Phi_{\mathbf{x}}(\mathbf{x},e^*))\in T^*M,\quad E(\mathbf{x})=0 \}. 
 \end{equation}
Note that we can equivalently use the thermodynamic potential with the difference between the energy and the total energy of the system, $E(\xx)-E_{\text{tot}}$, so the formulation using Morse family is not restricted to systems with zero energy. It is actually restricted to system with constant energy.

\textbf{GENERIC: Coupling Gradient and Hamiltonian Flows.}
Let us now merge a gradient flow and a Poisson flow defined on the state space $M$, by lifting them to the cotangent bundle level $T^*M$. For this, we define a Hamiltonian function on $T^*M$ as the sum of the negative of the dissipation potential and the function $F$ in \eqref{F} (generating the complete cotangent lift of the Hamiltonian flow),
\begin{equation}\label{Psi}
\Psi(\mathbf{x},\mathbf{x}^*)= \frac{1}{e^*}\mathbf{x}^{\ast }\cdot \mathbb{L} 
\Phi_{\mathbf{x}}-\Xi(\mathbf{x},\mathbf{x}^*),
\end{equation}
called the dynamic potential.
According to the local realization in \eqref{HamEqLoc}, the Hamiltonian function $\Psi(\mathbf{x},\mathbf{x}^*)$ on the cotangent bundle $T^*M$ determines a Hamiltonian vector field $X_\Psi$ on $T^*M$. A direct calculation proves that vector field $X_\Psi$ is simply the sum of minus of the Hamiltonian dynamics $X_\Xi$ generated by the dissipation potential (depicted in Diagram \ref{Xg}) and the
cotangent lift of Hamiltonian dynamics $X^{c*}$ in \eqref{X-c*},
\begin{equation}\label{X-Psi-}
X_\Psi=X^{c*} -X_\Xi.
\end{equation}

The exterior derivative $d\Phi$  of the thermodynamic potential  $\Phi$ in \eqref{Psi} is a section of the cotangent bundle. 
Then, referring to the commutative diagram 
\begin{equation}\label{Ham-GENERIC-----}
\xymatrix{ 
\mathbb{R}& M\times \mathbb{R} \ar[dd]^{\tau}\ar[l]_{\Phi}& T^{*}M\ar@(ul,ur)^{L_{\Phi}}\ar[dd]^{\pi_{M}} \ar[rrr]^{X_{\Psi}}& &
&TT^{*}M\ar[dd]_{T\pi_{M}} \\ & & &\\ &
M  \ar@{=}[r]& 
M \ar[rrr]^{X_{\Psi}^{d\Phi}}& & & TM}
\end{equation}
we define a projection of the Hamiltonian vector $X_\Psi$ to the base manifold $M$,
\begin{equation}\label{Generic-pre}
X_{\Psi}^{d\Phi}:= T\pi_{M} \circ X_{\Psi} \big \vert_{L_{\Phi}},
\end{equation}
which is the GENERIC flow. In terms of the local coordinates, 
GENERIC becomes
\begin{equation}
\frac{d \mathbf{x}}{dt} =\frac{1}{e^*}\mathbb{L}  \Phi_{\mathbf{x}} -\left.\Xi_{\mathbf{x}^*}
\right\vert_{\mathbf{x}^*= \Phi_\mathbf{x}},\qquad  \Phi_{e^*}=0.
\end{equation}

The complementary degeneracies (\ref{degg}) of the Poisson and the gradient structures allow us to rewrite GENERIC also as
\begin{equation}\label{Generic}
\frac{d \mathbf{x}}{dt} =\mathbb{L}  E_{\mathbf{x}} +  \left.\Xi_{\mathbf{x}^*}
\right\vert_{\mathbf{x}^*=S_\mathbf{x}}, \qquad  \Phi_{e^*}=0.
\end{equation}
In summary, GENERIC can be seen as the fiber projection of the vector field $X_\Psi$ restricted to the Lagrangian submanifold $L_\Phi$.
 
\textbf{GENERIC Flow vs. Hamiltonian Flow.} Let us now examine the relationship between the Hamiltonian dynamics $X_{\Psi}$ on $T^*M$, given in
\eqref{X-Psi-}, and GENERIC $
X_{\Psi}^{d\Phi}$ on $M$, given in Equation \eqref{Generic-pre}. The geometric Hamilton-Jacobi Theorem \ref{HJT} leads to the following statement.
\begin{proposition}\label{Prop-Ham}
The Hamiltonian dynamics $X_{\Psi}$ in
\eqref{HamEqLoc} and the GENERIC vector field $
X_{\Psi}^{d\Phi} $ in \eqref{grad} are related by
\begin{equation}
    Td\Phi \circ X_{\Psi}^{d\Phi}= X_{\Psi} \circ d\Phi \label{eq.compat-GENERIC}
\end{equation}
if and only if the dynamic potential is a solution of the stationary Hamilton-Jacobi equation, that is
\begin{equation}
\Psi(\mathbf{x} ,\Phi_\mathbf{x})=\epsilon
\end{equation}
for a constant $\epsilon$. 
\end{proposition}
The exterior derivative of the thermodynamic potential lifts integral curves to the GENERIC vector field $X_{\Xi}^{dS}$ to integral curves of the Hamiltonian vector field $X_{\Psi}$. In terms of the variational principles, the Hamilton-Jacobi equation is a result of criticality of the action integral
\begin{equation}
    \int_{t_0}^{t_1} \Psi^*(\mathbf{x},\dot{\mathbf{x}}) ~dt,
\end{equation}
with fixed initial time and point while keeping the final time and point undetermined. Here, $\Psi^*$  is the Lagrangian function obtained by the Legendre transformation of the Hamiltonian function $\Psi$. This action integral resembles the generalized Onsager-Machlup principle \cite{mielke-potential,om,renger-fluxes}.

\textbf{Pure Gradient Flow with Lagrange Multipliers.}
Another application of the Morse families is that gradient dynamics can be equipped with constrains in a geometric fashion. 
Consider an entropy $S=S\left( \mathbf{x}\right)$ on a manifold $M$ and some static
constraints  $\mathbf{w} ( 
\mathbf{x} )$, for instance the energy or total mass. Let us introduce a potential
\begin{equation*}
\Phi  ( \mathbf{x,w}^{\ast } ) =-S ( \mathbf{x} )
+\left\langle \mathbf{w}^{\ast },\mathbf{w} ( 
\mathbf{x} )
\right\rangle,
\end{equation*}
where $\mathbf{w}^{\ast }$ plays the role of Lagrange multiplier.
Potential $\Phi $ as a function on the total space of a
fibration $(Y,\tau ,M)$, where the total space admits local
coordinates $\left( \mathbf{x,w}^{\ast }\right)$, and it
determines a Morse family if the rank of $\mathbf{w}_{\mathbf{x}}$ is
maximal. Consequently, the Lagrangian submanifold determined by $\Phi$ in $T^{\ast }M$ is
\begin{equation*}
L_{\Phi}=\left\{  ( \mathbf{x,}\Phi _{\mathbf{x}} ) \in
T^{\ast }M :\mathbf{w} ( \mathbf{x} ) =0\right\} .
\end{equation*}%

The Hamiltonian on $T^*M$ is chosen as
\begin{equation*}
\Psi \left( \mathbf{x,x}^{\ast }\right) =-\Xi \left( \mathbf{x,x}^{\ast
}\right) 
\end{equation*}%
where $\Xi $ is a dissipation potential. Restriction  of the Hamiltonian vector
field $X_\Psi$ to the Lagrangian submanifold $L_{\Phi }$ then becomes
\begin{equation*}
X_{\Psi }\big\vert_{L_{\Phi }}=-\Xi _{\mathbf{x}^{\ast }}\big\vert%
_{\mathbf{x}^{\ast }=\Phi _{\mathbf{x}}}\cdot \nabla _{\mathbf{x}}+\Xi _{%
\mathbf{x}}\big\vert_{\mathbf{x}^{\ast }=\Phi _{\mathbf{x}}}\cdot \nabla _{%
\mathbf{x}^{\ast }},\qquad \mathbf{w} ( \mathbf{x} ) =0.
\end{equation*}%
Finally, the projection of this projection to the base manifold gives $T\pi _M\big(X _{\Psi }\big\vert_{R_{\Phi }}\big)$, which is the gradient flow subject to the constraint $w(\xx)=0$,
\begin{equation}\label{eq.Morse.w}
\frac{d\mathbf{x}}{dt}=-\Xi _{\mathbf{x}^{\ast }}\big\vert%
_{\mathbf{x}^{\ast }=\Phi _{\mathbf{x}}},\qquad\mathbf{w} ( \mathbf{x} ) =0.
\end{equation}
Note, however, that this dynamics is not generated by any explicitly specified vector field, but it represents rather a submanifold of $TM$, $T\pi _M\big(X _{\Psi }\big\vert_{L_{\Phi }}\big)$. Indeed, the derivative $\Xi_{\xx^*}$ contains also the Lagrange multipliers $w^*$, that are determined only implicitly by ensuring the constraint $w(\xx)=0$.
The construction is summarized in the following diagram:
\begin{equation}
\xymatrix{ \mathbb{R}& Y \ar[dd]^{\tau}\ar[l]_{\Phi}& T^{*}M  \ar@(ul,ur)^{L_{\Phi}} \ar[dd]^{\pi^0_M}
\ar[rrr]^{X_\Psi}& & &T T^{*}M \ar[dd]_{T\pi _M}\\ & & &\\ &
M  \ar@{=}[r]&M \ar[rrr]&
& & TM\ar@(dl,dr)_{T\pi _M\big(X _{\Psi}\big\vert_{R_{\Phi }}\big)} }  \label{Evo-HJ-5}
\end{equation}

\subsection{Some Examples of GENERIC}

We provide here two examples of the geometry presented in the previous section.

\textbf{Example: Morse family as dynamics with constraints.} 
Morse family is a geometric framework which makes it possible to write gradient dynamics with constraints. Let us now demonstrate it on a simple example where position in a plane $\xx=(x,y)$ evolves along the circle $0= w(\xx)=\Vert\xx\Vert^2-1$. Let the dissipation potential be $\Xi = (1/2) \zeta \Vert\xx^*\Vert^2$. The gradient dynamics \eqref{eq.Morse.w} then becomes 
 \begin{equation}
  \frac{d \xx}{dt} = \zeta S_{\xx} - 2\zeta w^* \xx ,
  \qquad w(\xx)=0.
 \end{equation}
 The condition that $w$ vanishes is satisfied when 
 \begin{equation}
  0 = \dot{w} = w_\xx \cdot \dot{\xx} = \zeta w_\xx \cdot S_\xx - 2 \zeta w^* (w_\xx)^2, 
 \end{equation}
 which determines the Lagrange multiplier $w^* = w_\xx\cdot S_\xx /(2(w_\xx)^2)$.
 In particular, the entropy can be chosen as $S(\xx) = -V(\xx)/T$, where $V(\xx)$ is a potential and $T$ a constant temperature, and the dynamics then becomes motion towards the minimum of potential $V(\xx)$ along the specified circle. In summary, gradient dynamics generated by the Morse family contains more terms than just the derivative of the dissipation potential with respect to the gradient of entropy, and the extra terms guarantee validity of the required constraints.

\textbf{Example: Conformal Hamiltonian Dynamics.}
This example contains another way towards GENERIC, at least a special case of it, within the geometric setting of conformal Hamiltonian dynamics.
Let us start with a base manifold $Q$, its cotangent bundle $T^*Q$, and the iterated cotangent bundle $T^*T^*Q$. In the Darboux coordinates $\mathbf{z}=(\mathbf{q},\mathbf{p})$ on $T^*Q$ and the Darboux' coordinates $(\mathbf{z},\boldsymbol{\Pi})=(\mathbf{q},\mathbf{p},\boldsymbol{\Pi}^q,\boldsymbol{\Pi}^p)$ on $T^*T^*Q$, the complete cotangent lift of the canonical Hamiltonian flow from $T^*Q$,
\begin{equation}
X_{H}=H_\mathbf{p}\cdot \nabla_\mathbf{q} -H_\mathbf{q}\cdot \nabla_\mathbf{p} \label{MV.Xham},
\end{equation}
to $T^*T^*Q$ reads
\begin{equation}
X_{H}^{c\ast }=X_{H}+(
\boldsymbol{\Pi}^p\cdot \nabla_\mathbf{q})H_\mathbf{q}\cdot \nabla_{\boldsymbol{\Pi}^q}-(
\boldsymbol{\Pi}^q\cdot \nabla_\mathbf{p})H_\mathbf{p}\cdot \nabla_{\boldsymbol{\Pi}^p}
.
\end{equation}
The lifted vector field is also Hamiltonian with respect to the canonical symplectic form $d\mathbf{q}\wedge d\boldsymbol{\Pi}^q +d\mathbf{p}\wedge d\boldsymbol{\Pi}^p$, and the corresponding Hamiltonian function is $\langle\boldsymbol{\Pi}, \mathbf{X}_H\rangle$, see \cite{esen2019lifts,EsGu12}. 

To add a dissipative term to the canonical Hamiltonian flow in \eqref{MV.Xham}, we introduce a dissipation potential $\Xi(\mathbf{q},\mathbf{p},\boldsymbol{\Pi}^q,\boldsymbol{\Pi}^p)$ on the iterated cotangent bundle $T^*T^*Q$. Let us choose the dissipation function as independent of $\boldsymbol{\Pi}^q$, that is
$\Xi=\Xi(\mathbf{q},\mathbf{p},\boldsymbol{\Pi}^p)$, in particular
\begin{equation}
\Psi(\mathbf{z},\boldsymbol{\Pi})= \boldsymbol{\Pi} \cdot \mathbf{X}_H - \Xi(\mathbf{q},\mathbf{p},\boldsymbol{\Pi}^p).
\end{equation}
This dissipation function determines also a corresponding Hamiltonian vector field $X_\Psi$ on $T^*T^*Q$. 
When we now choose an exact one-form $d\Phi$ on the base space $T^*Q$, we can define a projection $X_\Psi^{d\Phi}=T\pi_{T^*Q} \circ X_{\Psi}\circ  dH$, summarized in the following diagram,
\begin{equation}\label{Ham-GENERIC-T*Q}
\xymatrix{ T^{*}T^*Q\ar[dd]^{\pi_{T^*Q}} \ar[rrr]^{X_{\Psi}}& &
&TT^{*}T^*Q\ar[dd]_{T\pi_{T^*Q}} \\ & & &\\
T^*Q\ar@/^2pc/[uu]^{d\Phi}\ar[rrr]^{X_{\Psi}^{d\Phi}}& & & TT^*Q}
\end{equation}
where $\pi_{T^*Q}$ is the cotangent bundle projection from $T^*T^*Q$ to $T^*Q$.
Finally, the projected  dynamics $X_\Psi^{d\Phi}$ in the Darboux' coordinates becomes
\begin{equation}\label{gen-conf-dyn}
\dot{\mathbf{q}}=H_{\mathbf{p}},\qquad \dot{\mathbf{p}}=-H_{\mathbf{q}}+ \Xi_{\boldsymbol{\Pi}^p}\Big\vert_{\boldsymbol{\Pi}^p=\Phi_{\mathbf{p}}},
\end{equation}
which has the GENERIC structure.

Let us consider a particular case with dissipation potential $\Xi=c\Vert\boldsymbol{\Pi}^p\Vert^2$ (where $c$ is a constant), and potential $\Phi(\mathbf{p})=\Vert\mathbf{p}\Vert^2$. Equation \eqref{gen-conf-dyn} then leads to the following dynamics,
 \begin{equation}\label{gen-conf-dyn-}
\dot{\mathbf{q}}=H_{\mathbf{p}},\qquad \dot{\mathbf{p}}=-H_{\mathbf{q}}+ c\mathbf{p},
\end{equation}
called conformal Hamiltonian dynamics \cite{mclachlan2001conformal}. The conformal character of this system can be seen by taking the Lie derivative of the symplectic two-form $\Omega_Q$ (on $T^*Q$) with respect to vector field on the right hand side of Equation \eqref{gen-conf-dyn} (vector field $X_H^c$),
 \begin{equation}\label{prop-div-conf}
\mathcal{L}_{X_H^c}\Omega_Q=c\Omega_Q.
\end{equation}

\subsection{Rate GENERIC Dynamics} \label{rateGEN-VS}
Inspired by Extended Irreversible Thermodynamics, where fluxes of state variables become new variables, the dual variables become state variables within the rate-GENERIC \cite{jou-eit}. The dynamics of dual variables has often favorable mathematical properties like higher regularity or symmetric hyperbolicity \cite{godunov-interesting,god,godr,shtc-generic,diperna-lions}.

\textbf{Holonomic Lift of GENERIC Motion.} A geometric way towards dynamics of the dual variables starts with holonomic lifts of the GENERIC dynamics on a base manifold $M$. The lift gives a copy of the GENERIC flow \eqref{Generic} on the cotangent bundle $T^*M$.
According to definition \eqref{hol-lift-defn}, the holonomic lift of the GENERIC motion $X_{\Psi}^{d\Phi}$ is a generalized vector field
\begin{equation} \label{hol-Ham}
 X ^{hol}=\left. \Psi_{\mathbf{x}^*}\right\vert_{\mathbf{x}^*= \Phi_\mathbf{x}}\cdot \nabla_\mathbf{x} +\mathbb{U}
\left. \Psi_{\mathbf{x}^*}\right\vert_{\mathbf{x}^*= \Phi_\mathbf{x}}\cdot \nabla_{\mathbf{x}^*},
\end{equation}
where $\mathbb{U}=\mathbf{x}^*_\mathbf{x}=[ \partial x^*_i / \partial x^j ]$. 

Vector field $ X ^{hol}$ can be now restricted to the image space of the first jet of $\Phi=\Phi(\xx)$, $\mathbf{x}^*= \Phi_\mathbf{x}$ and $\mathbb{U}=\Phi_{\mathbf{xx}}$, which gives
\begin{equation} \label{idento}
 \frac{d \mathbf{x} }{dt} = \Psi_{\mathbf{x}^*}\big\vert_{\mathbf{x}^*= \Phi_\mathbf{x}}  ,\qquad 
 \frac{d \mathbf{x}^*}{dt} =\Phi_{\mathbf{xx}}  \Psi_{\mathbf{x}^*} \big\vert_{\mathbf{x}^*= \Phi_\mathbf{x}}.
\end{equation}

\textbf{Lifted Flow.} Diagram \ref{Ham-GENERIC} always commutes for the holonomic lift $X ^{hol}$ (replacing cotangent bundle $T^*M$ with the first jet bundle $J^1 \pi_M$ of the cotangent fibration), which is summarized in the following diagram,
\begin{equation}\label{Ham-GENERIC--}
\xymatrix{ J^1 \pi_M \ar[dd]^{\pi_{M}} \ar[rrr]^{X ^{hol}}& &
&TT^{*}M\ar[dd]_{T\pi_{M}} \\ & & &\\
M\ar@/^2pc/[uu]^{J\Phi}\ar[rrr]^{X_{\Psi}^{d\Phi}}& & & TM\ar@/_2pc/[uu]_{Td\Phi}}
\end{equation}
where $J^1\pi_M $ is the first manifold of the cotangent bundle, and $J\Phi$ is the jet prolongation of the function $\Phi$ given in \eqref{jet-sect}. 
This diagram shows, in particular, the identity
   \begin{equation}\label{dopo} 
   X ^{hol} \circ J\Phi = Td\Phi \circ X_{\Psi}^{d\Phi}.
   \end{equation}
Indeed, the left hand side of this identity is the dynamics in \eqref{idento}, whereas the right hand side is
    \begin{equation} 
   Td\Phi \circ X_{\Psi}^{d\Phi}=   \left. \Psi_{\mathbf{x}^*}\right\vert_{\mathbf{x}^*= \Phi_\mathbf{x}}\cdot \nabla_{\mathbf{x}} +
\Phi_{\mathbf{xx}} \left. \Psi_{\mathbf{x}^*}\right\vert_{\mathbf{x}^*= \Phi_\mathbf{x}} \cdot \nabla_{\mathbf{x}^*},
    \end{equation}
and identity \eqref{dopo} follows by direct comparison. Actually, the identity can be considered as an alternative definition of the holonomic lift. In other words, the holonomic lift of the GENERIC flow is precisely mimicking the dynamics on the base manifold, but this time the independent variables are both the position $\mathbf{x}$ and momenta $\mathbf{x}^*$. One can see this geometry as a way to copy the dynamics on the base manifold to the dual space.

\textbf{The Case of Linear Configuration Space.}
Let us now assume that $M$ is a vector space (not only a manifold), so that we can define the dual space $M^*$. The cotangent bundles of $M$ and $M^*$  are the same up to a reordering. In other words, there exits a bijection\footnote{This works in finite-dimensional spaces, while in the infinite-dimensional case, it becomes more involved \cite{roubicek}.}
\begin{equation}
\Gamma: T^{\ast }M=M\times M^{\ast } \longrightarrow T^{\ast }M^{\ast }=M^{\ast }\times M,\qquad (\mathbf{x},\mathbf{x}^*)\mapsto (\mathbf{x}^*,\mathbf{x}).
\end{equation}
The vector-space structure of $M$ allows two things: 

\textbf{(A)} We can define the Legendre transformation of the thermodynamic potential $\Phi=\Phi(\xx)$ denoted by $\Phi^*:M^*\mapsto \mathbb{R}$, which allows for the following commutation diagram,
\begin{equation}
\xymatrix{ T^{*}M=M\times M^* \ar[dd]^{\pi_{M}} \ar[rrr]^{\Gamma}& &
& T^{\ast }M^{\ast }=M^{\ast }\times M \ar[dd]_{\pi_{M^*}} \\ & & &\\
M\ar@/^2pc/[uu]^{d\Phi}\ar[rrr]^{Leg}& & & M^* \ar@/_2pc/[uu]_{d\Phi^*}}
\end{equation}
where $Leg$ is the induced Legendre transformation. The diagram expresses that, in particular,
\begin{equation}\label{Leg-dual}
\Gamma \circ d\Phi =  d\Phi^* \circ Leg, \qquad \mathbf{x}=\Phi^{\ast }_{\mathbf{x}^{\ast }}(\mathbf{x}^{\ast }),\quad 
\mathbf{x}^*=\Phi_{\mathbf{x}}(\mathbf{x}).
\end{equation}

We can also extend these relations to the jet manifold level by establishing a map from $J^1\pi_M$ to $J^1\pi_{M^*}$. First, we consider the following identifications
\begin{equation}
\mathbf{x}^*_{\mathbf{x}}=(\Phi_{\mathbf{x}})_\mathbf{x}=\Phi_{\mathbf{xx}},\qquad \mathbf{x}_{\mathbf{x}^*}=(\Phi^*_{\mathbf{x}^*})_{\mathbf{x}^*}=\Phi^*_{\mathbf{x}^*\mathbf{x}^*}
\end{equation}
and identities 
\begin{equation}\label{identities-I}
\mathbf{x}^*_{\mathbf{x}} \mathbf{x}_{\mathbf{x}^*}=\mathbb{I},\qquad \Phi_{\mathbf{xx}} \Phi^*_{\mathbf{x}^*\mathbf{x}^*}=\mathbb{I},
\end{equation}
where $\mathbb{I}$ is the identity matrix. The latter identity manifests that $\mathbb{U}=\Phi_{\mathbf{xx}}$ and $\mathbb{U}^{-1}=\Phi^*_{\mathbf{x}^*\mathbf{x}^*}$ are inverse matrices. Eventually, we have the following commutative diagram,
\begin{equation}\label{Jet-ident}
\xymatrix{ J^1\pi_M  \ar[dd]^{\pi_1} \ar[rrr]^{\hat{\Gamma}}& &
& J^1\pi_{M^*} \ar[dd]_{\pi_1} \\ & & &\\
M\ar@/^2pc/[uu]^{J \Phi}\ar[rrr]^{Leg}& & & M^* \ar@/_2pc/[uu]_{J \Phi^*}}
\end{equation}
where $J^1\pi_M$ is the first jet manifold for the fibration $\pi_M:T^*M\mapsto M$, whereas $J^1\pi_{M^*}$ is the first jet manifold for $\pi_{M^*}:T^*M^*\mapsto M^*$. Notice that $J \Phi$ and $J \Phi^*$ are the first prolongations (see \eqref{jet-sect} for the definition) of the functions $\Phi$ and $\Phi^*$, respectively. In local coordinates they are given by
\begin{equation}
J \Phi(\mathbf{x})=(\mathbf{x},\Phi_\mathbf{x},\Phi_\mathbf{xx}),\qquad 
J \Phi^*(\mathbf{x}^*)=(\mathbf{x}^*,\Phi_{\mathbf{x}^*},\Phi_{\mathbf{x}^*\mathbf{x}^*}).
\end{equation}

\textbf{(B)} Our next task is to find dynamics of on $M^*$ (dynamics of thermodynamic forces). Referring to Diagram \ref{Jet-ident}, we  substitute the jet bundle $J^1\pi_M$ with the jet bundle $J^1\pi_{M^*}$ in the definition of the holonomic lift $\big(X_{\Psi}^{d\Phi}\big)^{hol}$ given in \eqref{hol-Ham} of the GENERIC flow. Then, we can see $\big(X_{\Psi}^{d\Phi}\big)^{hol}$ as a generalized vector field on the dual space $M^*$, which becomes in the local coordinates,
\begin{equation} \label{hol-Ham-dual}
X^{hol}=\left. \Psi_{\mathbf{x}^*}\right\vert_{\mathbf{x}^*= \Phi_\mathbf{x}}\cdot \nabla_\mathbf{x} + [\Phi^*_{\mathbf{x}^*\mathbf{x}^*}]^{-1} 
\left. \Psi_{\mathbf{x}^*}\right\vert_{\mathbf{x}^*= \Phi_\mathbf{x}}\cdot\nabla_{\mathbf{x}^*},
\end{equation} 
where $[\Phi^*_{\mathbf{x}^*\mathbf{x}^*}]^{-1} $ is the inverse of the Hessian matrix $\Phi^*_{\mathbf{x}^*\mathbf{x}^*}$. 

Projection of $\big(X_{\Psi}^{d\Phi}\big)^{hol}$ to the base manifold $M^*$ by means of the section $J^1\Phi^*$ gives 
\begin{equation}
\big(X^{hol}\big)^{J\Phi^*}:=T\pi _{M^*} \circ X^{hol} \circ J\Phi^*,
\end{equation} 
which is depicted in the following diagram,
\begin{equation}\label{Diag-Y}
\xymatrix{  J^1\pi_{M^*}
\ar[dd]^{\pi_{1}} \ar[rrr]^{X^{hol}}&   & &T T^{*}M^* \ar[dd]^{T\pi _{M^*}}\\
  &  & &\\
M ^{*} \ar@/^2pc/[uu]^{J\Phi^*}\ar[rrr]^{\big(X^{hol}\big)^{J\Phi^*}}&  & & TM^{*} }
\end{equation} 
where $\pi_{M^*}$ is the cotangent bundle projection from $T^{*}M ^{*}$ to $M^*$. 
In local coordinates $(\mathbf{x} ^*,\mathbf{x} )$ on $ T^{*}M ^{*}$, the projected dynamics, called rate GENERIC, becomes
\begin{equation}\label{Rate-GENE}
\frac{d \mathbf{x}^*}{dt}=[\Phi^*_{\mathbf{x}^*\mathbf{x}^*}]^{-1} 
\left. \Psi_{\mathbf{x}^*}\right \vert_{\mathbf{x}= \Phi^*_{\mathbf{x}^*}(\mathbf{x}^*)} .
\end{equation} 
Rate GENERIC will be further discussed in follow-up paper \cite{OMM-2}.
Notice that the holonomic lift of the GENERIC flow given in \eqref{hol-Ham-dual} determines both GENERIC and rate GENERIC dynamics in a collective form. 

\textbf{The Legendre Transformation.}
Recalling the Legendre transformation, we can now examine the equivalence between GENERIC and  rate-GENERIC flows. The tangent mapping of the Legendre transformation commutes as in the following diagram,
\begin{equation}
\xymatrix{ TM  \ar[dd]^{\tau_{M}} \ar[rrr]^{TLeg }& &
& T M^{\ast }  \ar[dd]_{\tau_{M^*}} \\ & & &\\
M\ar@/^2pc/[uu]^{X^{d\Phi}_\Psi}\ar[rrr]^{Leg}& & & M^* \ar@/_2pc/[uu]_{Leg_{*}X^{d\Phi}_\Psi }}
\end{equation}
In terms of the induced coordinates $(\mathbf{x},\dot{\mathbf{x}})$ on $TM$ and the induced coordinates on $(\mathbf{x}^*_i,\dot{\mathbf{x}}^*)$, 
we have that
\begin{equation}
TLeg: TM \longrightarrow T M^{\ast }, \qquad (\mathbf{x},\dot{\mathbf{x}})
\mapsto (\Phi_{\mathbf{x}} , \Phi_{\mathbf{xx}} \dot{\mathbf{x}}).
\end{equation}
Therefore, one maps the GENERIC flow on $M$ to the rate GENERIC flow on $M^*$ as 
\begin{equation}\label{h-gene}
\frac{d \mathbf{x}^*}{dt} =
 \Phi_{\mathbf{xx}}\left. \frac{d \mathbf{x}}{dt} \right\vert_{\mathbf{x}^*= \Phi_\mathbf{x}}
 =\Phi_{\mathbf{xx}} \left. \Psi_{\mathbf{x}^*}\right\vert_{\mathbf{x}^*= \Phi_\mathbf{x}}.
\end{equation}
By application of the second identity in \eqref{identities-I} on Equation \eqref{h-gene}, we arrive at the rate GENERIC dynamics in \eqref{Rate-GENE}. 

\textbf{Indefinite Form of Rate GENERIC.} The convexity of the thermodynamic potential $\Phi$ in the rate GENERIC \eqref{Rate-GENE} manifests in the positive semi-definiteness of Hessian matrix  $[\Phi^*_{\mathbf{x}^*\mathbf{x}^*}]^{-1} (\mathbf{x}^*)$.
The rate GENERIC equations can be generalized by replacing $[\Phi^*_{\mathbf{x}^*\mathbf{x}^*}]^{-1} (\mathbf{x}^*)$ by an arbitrary semi-positive definite metric $\mathbb{G}(\mathbf{x}^*)=[G_{ij}(\mathbf{x}^*)]$, while considering the function $\Psi=\Psi(\mathbf{x},\mathbf{x}^*,\mathbf{y}^{\dag})$ as unspecified. The scalars $\mathbf{y}^{\dag}$ then determine a set of Lagrange multipliers, and
rate GENERIC acquires the following indefinite form,
\begin{equation}\label{r-GENERIC}
\frac{d \mathbf{x}^*}{dt} = \mathbb{G}(\mathbf{x}^*)  \Psi_{\mathbf{x}^*},
\end{equation}
governing the time evolution of $\mathbf{x}^*$. 

It can be verified by direct calculation that (\ref{r-GENERIC}) together with the rate constitutive relations
\begin{equation}\label{GG}
\begin{split}
\mathbb{G}&=[\Phi^{\ast }_{\mathbf{x}^* \mathbf{x}^*} ]^{-1}\\ 
\Psi(\mathbf{x},\mathbf{x}^*,\mathbf{y}^{\dag})&=-\Xi(\mathbf{x},\mathbf{x}^*)+\langle \mathbf{y}(\mathbf{x}^*),\mathbf{y}^{\dag}\rangle\\ 
\mathbf{x} &=\Phi^*_{\mathbf{x}^*}(\mathbf{x}^*)\\ 
\mathbf{y}(\mathbf{x}^*)&=\mathbf{x}^*\\ 
\mathbf{y}^{\dag}&=e^*\mathbb{L}\Phi_\mathbf{x}(\mathbf{x})
\end{split}
\end{equation}
turn out to be the rate GENERIC equation in the form of \eqref{Rate-GENE} (in particular, it becomes GENERIC (\ref{Generic}) as well). The two equations  (\ref{Generic}) and (\ref{r-GENERIC}) are actually the same equations, related by the one-to-one transformation in \eqref{Leg-dual}.

\subsection{Rate GENERIC as Lift to the Iterated Cotangent Bundle}
In this Section, we write the rate Generic equation in \eqref{h-gene} in terms of the HJ theory. For simplicity, we consider only the dissipative dynamics without any Hamiltonian part. 

Let us start with a manifold $M$ with coordinates $\mathbf{x}$ and the cotangent bundle $%
T^{\ast }M$ with coordinates $\left( \mathbf{x},\mathbf{x}^{\ast }\right) $.
The iterated cotangent bundle $T^{\ast }T^{\ast }M=T^{\ast }\left( T^{\ast
}M\right) $ admits Darboux' coordinates, $\left( \mathbf{x},%
\mathbf{x}^{\ast };\mathbf{x}^{\top },\mathbf{x}^{\dag }\right) $, and it is equipped with the symplectic two-form $\omega=d\mathbf{x}\wedge d\mathbf{x}^{\top }+d\mathbf{x}^{\ast }\wedge \mathbf{x}^{\dag }$. For a Hamiltonian function $H=H ( \mathbf{x},\mathbf{x}^{\ast };\mathbf{x}^{\top }\mathbf{,x}^{\dag })$, the Hamiltonian vector field
on  $T^{\ast }T^{\ast }M $ is 
\begin{equation}
X _{H}=H_{\mathbf{x}^{\top }}\cdot \nabla _{\mathbf{x}}+H_{\mathbf{%
x}^{\dag }}\cdot \nabla _{\mathbf{x}^{\ast }}- H_{\mathbf{x}} \cdot \nabla _{\mathbf{x}^{\top }}- H_{\mathbf{x}^{\ast }} \cdot \nabla _{\mathbf{x}^{\dag }}.
\end{equation}%

Moreover, we consider
the following functions,
\begin{equation}\label{fcts}
\begin{split}
\Phi &: M\rightarrow  
\mathbb{R} 
,\qquad \mathbf{x}\rightarrow -S\left( \mathbf{x}\right) \\
\Psi &: T^{\ast }M\rightarrow  
\mathbb{R} 
,\qquad\left( \mathbf{x},\mathbf{x}^{\ast }\right) \rightarrow -\Xi
\left( \mathbf{x,x}^{\ast }\right) \\
\Omega &: T^{\ast }T^{\ast }M\rightarrow  
\mathbb{R} 
,\qquad\left( \mathbf{x},\mathbf{x}^{\ast };\mathbf{x}^{\top }\mathbf{%
,x}^{\dag }\right) \rightarrow (1/2) \langle \mathbf{x}^{\dag },S_{%
\mathbf{xx}}\mathbf{x}^{\dag } \rangle ,
\end{split}
\end{equation}
called static potential, dynamic
potential, and dynamic super potential, 
respectively. Here, $S_{\mathbf{xx}}$ stands for the Hessian matrix of the entropy $%
S $. 
The dynamic super potential $\Omega $ can be chosen as the Hamiltonian function. It is independent of
the covector variables $\mathbf{x}^{*}$ and $\mathbf{x}^{\top }$, and it is a
quadratic function of $\mathbf{x}^{\dag }$. 
The corresponding Hamiltonian vector field is then
\begin{equation}\label{Ham-omega}
X_{\Omega }=S_{\mathbf{xx}}\mathbf{x}^{\dag }\cdot \nabla _{%
\mathbf{x}^{\ast }}-\frac{1}{2}\left\langle \mathbf{x}^{\dag },S_{\mathbf{xx}%
}\mathbf{x}^{\dag }\right\rangle _{\mathbf{x}}\cdot \nabla _{\mathbf{x} 
^{\top }},
\end{equation}%
whereas the nonzero dynamical equations are%
\begin{equation*}
\frac{d\mathbf{x}^{\ast }}{dt}=S_{\mathbf{xx}}\mathbf{x}^{\dag },\qquad 
\frac{d\mathbf{x}^{\top }}{dt}=-\frac{1}{2}\left\langle \mathbf{x}^{\dag
},S_{\mathbf{xx}}\mathbf{x}^{\dag }\right\rangle _{\mathbf{x}}.
\end{equation*}%

Let us now examine the dynamics restricted to the image of the exterior derivative $d\Psi$ of the dynamical potential,
\begin{equation*}
X_{H}\circ d\Psi =S_{\mathbf{xx}}\Psi _{\mathbf{%
x}^{\ast }}\cdot \nabla _{\mathbf{x}^{\ast }}-\frac{1}{2}\left\langle \Psi _{ 
\mathbf{x}^{\ast }},S_{\mathbf{xxx}}\Psi _{\mathbf{x}^{\ast }}\right\rangle  \cdot \nabla _{\mathbf{x}^{\top }}.
\end{equation*}%
By projecting the vector field to the base manifold $T^{\ast }M$, we have the
projected vector field $X_{\Omega}^{d\Psi }$, as can be seen in the following diagram,
\begin{equation}
\xymatrix{  T^{*}T^{*}M  \ar[dd]^{\pi _{T^{*}M}}
\ar[rrr]^{X_\Omega}& & &T T^{*}T^{*}M \ar[dd]^{T\pi _{T^{\ast }M}} \\ & & &\\ T^*M \ar@/^2pc/[uu]^{d\Psi} \ar[rrr]^{X _{\Omega}^{d\Psi }}&
& & TT^*M  }  \label{Evo-HJ-2}
\end{equation}
Explicitly, we have that
\begin{equation*}
X_{\Omega}^{d\Psi }:=T\pi _{T^{\ast }M}\circ
X _{\Omega}\circ d\Psi (\mathbf{x},\mathbf{x}^*) =S_{\mathbf{xx}}\Psi _{\mathbf{%
x}^{\ast }}\cdot \nabla _{\mathbf{x}^{\ast }}
\end{equation*}%
and the nonzero terms in the dynamics are
\begin{equation*}
\frac{d\mathbf{x}^{\ast }}{dt}=S_{\mathbf{xx}}\Psi _{\mathbf{x}^{\ast }}.
\end{equation*}
Note that $\xx$ turns out to be a parameter in this case. 

\textbf{Dynamical Potential with Lagrange multipliers.} Finally, the rate GENERIC can be also equipped with constraints. Consider fibration $Y^{\dag }\rightarrow T^{\ast }M$ and fiber coordinates $(\mathbf{x},\mathbf{x}^*,\mathbf{y}^{\dag })$. Function
\begin{equation*}
\Psi :Y^{\dag }\longrightarrow  
\mathbb{R} ,\qquad \left( \mathbf{x},\mathbf{x}^{\ast },\mathbf{y}^{\dag }\right)
\mapsto -\Xi \left( \mathbf{x,x}^{\ast }\right) +\left\langle \mathbf{y}^{\dag },\mathbf{y} ( \mathbf{x} ) \right\rangle
\end{equation*}
defines a Morse family if the rank of
the matrix
\begin{equation*}
(\Psi_{\mathbf{x}\mathbf{y}^{\dag }} \quad \Psi_{\mathbf{x}^*\mathbf{y}^{\dag }}\quad \Psi_{\mathbf{y}^{\dag }\mathbf{y}^{\dag }})= (\mathbf{y}_{\mathbf{x}} \quad\mathbf{ 0} \quad \mathbf{ 0})
\end{equation*}
is maximal, that is if the rank of $\mathbf{y}_{\mathbf{x}}$ is maximal. When we have the Morse family, we can consider
the Lagrangian submanifold determined by $\Psi $,
\begin{equation*}
L_{\Psi }=\left\{ \left( \mathbf{x},\mathbf{x}^{\ast },\Psi _{\mathbf{x}%
}\left( \mathbf{x},\mathbf{x}^{\ast },\mathbf{y}^{\dag }\right) ,\Psi _{%
\mathbf{x}^{\ast }}\left( \mathbf{x},\mathbf{x}^{\ast },\mathbf{y}^{\dag
}\right) \right) \in T^{\ast }T^{\ast }M:\mathbf{y}\left( \mathbf{x}\right)
=0\right\} .
\end{equation*}%
The restriction of the Hamiltonian vector field $X_{\Omega}$ to this Lagrangian submanifold is 
\begin{equation*}
X_{\Omega }\big\vert_{L_{\Psi }}=S_{\mathbf{xx}}\Psi _{\mathbf{x}%
^{\ast }}\cdot \nabla _{\mathbf{x}^{\ast }}-\frac{1}{2}\left\langle \Psi _{%
\mathbf{x}^{\ast }},S_{\mathbf{xxx}}\Psi _{\mathbf{x}^{\ast }}\right\rangle  \cdot \nabla _{\mathbf{x}^{\top }} ,\qquad \mathbf{y}\left( \mathbf{x}\right) =0,
\end{equation*}%
and the nonzero terms in the dynamical equations become
\begin{equation*}
\frac{d\mathbf{x}^{\ast }}{dt}=S_{\mathbf{xx}}\Psi _{\mathbf{x}^{\ast
}},\qquad \frac{d\mathbf{x}^{\top }}{dt}=-\frac{1}{2}\left\langle \Psi _{%
\mathbf{x}^{\ast }},S_{\mathbf{xxx}}\Psi _{\mathbf{x}^{\ast }}\right\rangle
,\qquad \mathbf{y}\left( \mathbf{x}%
\right) =0.
\end{equation*}
This is summarized in the following diagram: 
\begin{equation}
\xymatrix{ \mathbb{R}& Y^{\dag} \ar[dd]^{\tau}\ar[l]_{\Psi}& \ar@(ul,ur)^{L_{\Psi}}  T^{*}T^{*}M  \ar[dd]^{\pi _{T^{*}M}}
\ar[rrr]^{X_\Omega}& & &T T^{*}T^{*}M \ar[dd]^{T\pi _{T^{*}M}} \\ & & &\\ &
T^{\ast }M \ar@{=}[r]& T^*M   \ar[rrr]^{X _{\Omega}^{d\Psi }}&
& & TT^*M  \ar@(dl,dr)_{T\pi _{T^{*}M}(X _{\Omega }\big\vert_{L_{\Psi }})} }  \label{Evo-HJ-3}
\end{equation} 
By projecting $X_{\Omega }\big\vert_{L_{\Psi }}$ to the base manifold $T^{\ast }M^{\ast }$ with $T\pi _{T^{*}M}$, we arrive
at an implicit dynamics $T\pi _{T^{*}M}(X _{\Omega }\big\vert_{R_{\Psi }})$,
\begin{equation*}
\frac{d\mathbf{x}^{\ast }}{dt}=S_{\mathbf{xx}}\Psi _{\mathbf{x}^{\ast
}},\qquad \mathbf{y}\left( \mathbf{x}\right) =0,
\end{equation*}
satisfying the constraints. Once again, $\mathbf{x}$ turns out to be a set of parameters. 

\section{On GENERIC in Contact Geometry}\label{4}
While in the preceding sections we work with cotangent bundles, which are even-dimensional, in the current Section we step into contact geometry, which is odd-dimensional. The purpose is to promote thermodynamic potentials to state variables, which makes the second law of thermodynamics directly visible. 

\subsection{Hamiltonian Dynamics on Contact Manifold}\label{Con-Man-Sec}
A $(2n+1)-$dimensional manifold is called a contact manifold if it admits a contact one-form $\eta$ satisfying $d\eta^n
\wedge \eta \neq 0$,  \cite{Arnold-book,Liber87}. For a contact manifold, there exists a distinguished vector field called Reeb field, $\mathcal{R}$,  satisfying 
\begin{equation}
\iota_{\mathcal{R}}\eta =1,\qquad \iota_{\mathcal{R}}d\eta =0,
\end{equation}
where $\iota$ is the contraction mapping called interior derivative. 
The kernel $\ker \eta$ of the contact form $\eta$ at each point of the contact manifold determines the contact structure. 

\textbf{Contact Hamiltonian Dynamics.} 
For a real-valued function $H$ on the contact manifold $(M,\eta)$, the contact Hamiltonian vector field $X_H$ is defined as \cite{Br17,BrCrTa17,LeLa19}
\begin{equation}
\iota_{X_{H}}\eta =-H,\qquad \iota_{X_{H}}d\eta  =dH-\mathcal{R}(H) \eta,   \label{contact}
\end{equation}%
where $\mathcal{R}$ is the Reeb vector field. The Lie derivative of the contact one-form $\eta$ along a contact Hamiltonian vector field is
\begin{equation}\label{L-X-eta}
\mathcal{L}_{X _{H}}\eta =
d\iota_{X _{H}}\eta+\iota_{X _{H}}d\eta= -\mathcal{R}(H)\eta.
\end{equation}
In other words, the flow of a contact Hamiltonian system preserves the contact structure, but not the contact one-form. As a manifestation of \eqref{L-X-eta}, the Hamiltonian motion does not preserve the volume form $d\eta^n
\wedge \eta$, since
\begin{equation} 
{\mathcal{L}}_{X _H}  \, (d\eta^n
\wedge \eta) = - (n+1)  \mathcal{R}(H) d\eta^n
\wedge \eta.
\end{equation}%
Moreover, neither the Hamiltonian function is preserved along the motion,
\begin{equation} \label{L-X-H}
{\mathcal{L}}_{X _H} \, H = - \mathcal{R}(H) H,
\end{equation}
if not equal to zero.

\textbf{Self-consistency of Contact Hamiltonian Dynamics.}
A feature of Poisson geometry, where the Hamiltonian vector field is generated from a Poisson bivector and energy, is that neither the bivector nor energy vary along the evolution. More precisely, the Lie derivative of the Poisson bivector along the Hamiltonian vector field, as well as the Lie derivative of energy, are zero. Since both the contact one-form and the Hamiltonian vary along the integral curves of the contact Hamiltonian vector field, contact geometry is self-consistent only in a weaker sense. 

The contact Hamiltonian vector field is defined by vanishing $\iota_{X_H} \eta+H=0$. Lie derivative of this expression is
\begin{equation}\label{cons-cont}
  \mathcal{L}_{X_H} \left(\iota_{X_H} \eta+H\right) = \iota_{X_H} \mathcal{L}_{X_H}\eta + \mathcal{L}_{X_H}H
  =-\mathcal{R}(H)\iota_{X_H}\eta - \mathcal{R}(H)H 
  = 0,
\end{equation}
which means that this building block of the definition is indeed preserved by the contact Hamiltonian dynamics. Note that the Lie derivative commutes with the interior derivative, $\mathcal{L}_{X}\circ \iota_X=\iota_X \circ \mathcal{L}_{X}$, and that in the calculation of \eqref{cons-cont}, we have used identities \eqref{L-X-eta} and \eqref{L-X-H}. 

Similarly, the Lie derivative of the second part of the definition of the contact Hamiltonian vector field reads
\begin{equation}
\begin{split}
&\mathcal{L}_{X_H}\left(\iota_{X_H}d\eta -dH +\mathcal{R}(H)\eta\right)=\mathcal{L}_{X_H}\iota_{X_H}d\eta - \mathcal{L}_{X_H} (dH) + \mathcal{L}_{X_H}\big(\mathcal{R}(H)\eta \big)\\&\qquad \qquad
= \iota_{X_H} d \mathcal{L}_{X_H}\eta -d\mathcal{L}_{X_H}H +\mathcal{L}_{X_H}\big(\mathcal{R}(H)\eta \big)
\\&\qquad \qquad = -
\iota_{X_H} d \big(\mathcal{R}(H)\eta \big)  +d\big(\mathcal{R}(H) H \big) +d\iota_{X_H}\big(\mathcal{R}(H)\eta \big) + \iota_{X_H}d\big(\mathcal{R}(H)\eta \big)
\\&\qquad \qquad = d\big(\mathcal{R}(H) H \big)  - d\big(\mathcal{R}(H)H \big) =0.
\end{split}
\end{equation}
Therefore, contact Hamiltonian dynamics is self-consistent in the sense that it preserves its definition.

\textbf{Extended Cotangent Bundle.}
Consider an $n$-dimensional manifold $M$. Its 
extended cotangent bundle $T^*M\times \mathbb{R}$ is a $(2n+1)-$dimensional contact manifold, and in the  Darboux' coordinates, $(\mathbf{x},{\mathbf{x}^*},z)$, on $T^*M\times \mathbb{R}$, the contact one-form and the Reeb vector field become
\begin{equation}
\eta = d z -  \mathbf{x}^* \cdot  d \mathbf{x}, \qquad \mathcal{R}=\nabla_z,
\end{equation}
respectively. 
For a given Hamiltonian function $H$, the Hamiltonian vector field \eqref{contact} becomes
\begin{equation}\label{con-dyn}
X _H=H_{\mathbf{x}^*} \cdot \nabla_\mathbf{x}
-(H_\mathbf{x}+{\mathbf{x}^*}H_z)\cdot \nabla_{\mathbf{x}^*}
+({\mathbf{x}^*}\cdot H_{\mathbf{x}^*} - H)\nabla_z,
\end{equation}
and the contact Hamilton's equations are
\begin{equation}\label{conham}
	\frac{d \mathbf{x}}{dt}= H_{\mathbf{x}^*}, \qquad \frac{d {\mathbf{x}^*}}{dt}  = -H_\mathbf{x}+{\mathbf{x}^*}H_z, \qquad \frac{d z}{dt} = {\mathbf{x}^*}\cdot H_{\mathbf{x}^*} - H.
\end{equation}

\textbf{Evolution Hamiltonian Dynamics.} Unlike Hamiltonian mechanics, where the Hamiltonian vector field is the only geometrically distinguished vector field on the cotangent bundle, contact geometry admits two alternatives. In the preceding Section, we have recalled the contact Hamiltonian vector field $X_H$, while in the current Section we discuss the evolution Hamiltonian vector field $\varepsilon_H$ \cite{esen2021contact,simoes2020geometry,simoes2020contact}, determined through the following equalities,
\begin{equation}\label{evo-def} 
\iota_{\varepsilon_{H}}\eta =0,\qquad \iota_{\varepsilon_{H}}d\eta=dH-\mathcal{R}(H) \eta.
\end{equation} 

The Lie derivative of the contact one-form along the evolution vector field is
\begin{equation}\label{L-E-eta}
\mathcal{L}_{\varepsilon _{H}}\eta =
d\iota_{\varepsilon _{H}}\eta+\iota_{\varepsilon _{H}}d\eta= dH-\mathcal{R}(H) \eta.
\end{equation} 
Therefore, $\varepsilon_H$ does not preserve the contact structure, while the contact vector field $X_{H}$ does.

When we take the interior derivative of the second identity in \eqref{evo-def} with respect to $\iota_{\varepsilon _{H}}$, we arrive at
\begin{equation}\label{L-E-H}
\mathcal{L}_{\varepsilon _{H}}H=0, 
\end{equation}
which means that the evolution vector field $\varepsilon_H$ preserves the Hamiltonian function $H$. Note that the contact Hamiltonian vector field $X_{H}$ does not preserve the Hamiltonian.

In the Darboux' coordinates, the evolution Hamiltonian vector field becomes
\begin{equation}\label{evo-dyn}
	\varepsilon_H=H_{\mathbf{x}^*} \cdot \nabla_\mathbf{x}
-(H_\mathbf{x}+{\mathbf{x}^*}H_z)\cdot \nabla_{\mathbf{x}^*}
+({\mathbf{x}^*}\cdot H_{\mathbf{x}^*})\nabla_z,
\end{equation}
and the corresponding evolution equations are 
\begin{equation}\label{evo-eq}
\frac{d \mathbf{x}}{dt}= H_{\mathbf{x}^*}, \qquad \frac{d {\mathbf{x}^*}}{dt}  = -H_\mathbf{x}-{\mathbf{x}^*}H_z, \qquad \frac{d z}{dt} = {\mathbf{x}^*}\cdot H_{\mathbf{x}^*}.
\end{equation}
The difference between the evolution Hamiltonian flow $\varepsilon_H$ in \eqref{evo-dyn} and the contact Hamiltonian dynamics $X_H$ in \eqref{con-dyn} is the missing $-H$ term in $\varepsilon_H$ in the basis $\nabla_z$.


\textbf{A Simple Example.}
Consider the Darboux' coordinates $(\qq,\pp,S)$ and Hamiltonian $H = \Vert\pp\Vert^2/{2m} + V(\qq) + \zeta S$. Then the dynamics generated by the evolution vector field $\varepsilon_H$ reads
\begin{equation}
\frac{d \qq}{dt} =  \frac{\pp}{m},\qquad 
\frac{d \pp} {dt}= -V_\qq - \frac{\zeta \pp}{m} ,\qquad 
\frac{d S} {dt}  =  \frac{1}{m}\Vert\pp\Vert^2\geq 0.
\end{equation}
The first equation is the usual relation between velocity and momentum, the second equations represents Newton's law with linear friction, and the third equation tells that entropy is produced (the second law of thermodynamics). From the physical point of view, however, this formulation of a particle with friction is problematic because the energy should not depend on the friction coefficient and because the friction coefficient should affect the entropy production. This deficiency will be removed later in Section \ref{sec.Gen.Evo}.

\textbf{Self-consistency of Evolution Hamiltonian Dynamics}. 
Similarly as the contact Hamiltonian dynamics, the evolution Hamiltonian dynamics is self-consistent in the sense that the Lie derivative of the building blocks of the evolutionary vector field is zero. Indeed, we have that 
\begin{equation}
\mathcal{L}_{\varepsilon_H} \iota_{\varepsilon_H} \eta =\iota_{\varepsilon_H} \iota_{\varepsilon_H}d \eta =0
\end{equation}
due to the skew-symmetry of the two-form $d \eta$. On the other hand, we compute 
\begin{equation}
\begin{split}
&\mathcal{L}_{\varepsilon_H} (\iota_{\varepsilon_H}d\eta -dH + \mathcal{R}(H)\eta)= \mathcal{L}_{\varepsilon_H} \iota_{\varepsilon_H}d\eta -\mathcal{L}_{\varepsilon_H} dH + \mathcal{L}_{\varepsilon_H} \big(\mathcal{R}(H)\eta \big)
\\& \qquad \qquad =
\iota_{\varepsilon_H}d\mathcal{L}_{\varepsilon_H} \eta -d\mathcal{L}_{\varepsilon_H} H + \iota_{\varepsilon_H}d \big(\mathcal{R}(H)\eta \big)
+d
\iota_{\varepsilon_H} \big(\mathcal{R}(H)\eta \big)
\\& \qquad \qquad =\iota_{\varepsilon_H}d(dH-\mathcal{R}(H)\eta) + \iota_{\varepsilon_H}d \big(\mathcal{R}(H)\eta \big)
+d \big(\mathcal{R}(H)
\iota_{\varepsilon_H}\eta
\big)=0,
\end{split}
\end{equation}
where we have employed the identities \eqref{L-E-eta} and \eqref{L-E-H}.

\subsection{Geometric Hamilton-Jacobi Theories in Contact Geometry} 
Our goal is now to transform the preceding results on Hamilton-Jacobi theory and GENERIC into contact geometry. 
HJ Theory has been discussed in the context of contact geometry from various perspectives \cite{cannarsa2019herglotz,
de2021hamilton,de2017cosymplectic,
esen2021implicit,grillo2020extended}.
First, we consider the trivial line bundle $M\times \mathbb{R}\mapsto M$ over a manifold $M$. The first jet bundle is precisely equal to the extended cotangent bundle $T^*M\times \mathbb{R}$ with the projection
\begin{equation}
\begin{split}
\pi^0_M&:T^*M\times \mathbb{R}\longrightarrow M, \qquad  (\mathbf{x},{\mathbf{x}^*},z)\mapsto  \mathbf{x}.
\end{split}
\end{equation}
The first prolongation of a real-valued function $W$ on $M$ to the extended cotangent bundle $T^*M\times \mathbb{R}$ is
\begin{equation}\label{j1F}
\mathcal{T}^* W:M\longrightarrow  T^*M\times \mathbb{R},\qquad (\mathbf{x})\mapsto (\mathbf{x},W_{\mathbf{x}},W).
\end{equation}

\textbf{HJ for Evolution Dynamics.} Consider the evolution vector field $\varepsilon_H$, defined in Equation \eqref{evo-dyn}, for a Hamiltonian function $H$ (we do not require the condition $\mathcal{R}(H)=0$ here). Vector field $\varepsilon_H$ can be projected to the base manifold $M$, which gives a vector field on $M$,
 \begin{equation}\label{varep-H-proj}
\varepsilon^{\mathcal{T}^*W}_H:= T\pi^0_M \circ  X_H \circ \mathcal{T}^*W.
\end{equation}
Finally, we can formulate a version of the Hamilton-Jacobi theorem for the evolutionary vector field \cite{esen2021implicit}.
\begin{theorem} \label{HJT-Evo-Con-1}
For a smooth function $W=W(\xx)$ on $M$, the following conditions are equivalent:
\begin{enumerate}
\item The vector fields $\varepsilon_{H}$ and $\varepsilon^{\mathcal{T}^*W}_H$ are $\mathcal{T}^*W$-related, that is
\begin{equation}\label{HJT-Con-1-eq-1}
T\mathcal{T}^*W \circ  \varepsilon_H^{\mathcal{T}^*W} =\varepsilon_H\circ \mathcal{T}^*W,
\end{equation}
where $T\mathcal{T}^*W:TM\mapsto T(T^{*}M\times \mathbb{R})$ is tangent mapping of the prolongation $\mathcal{T}^*W$. 
\item The following exterior derivative is zero,
\begin{equation}\label{HJT-Con-1-eq-2-}
d( H\circ \mathcal{T}^*W  )=0.
\end{equation}
\end{enumerate}
\end{theorem}

In terms of the Darboux' coordinates, the Hamilton-Jacobi equation \eqref{HJT-Con-1-eq-2-} becomes
 \begin{equation}\label{evo-con-HJ}
H (\mathbf{x},W_{\mathbf{x}},W)=\epsilon,
 \end{equation}
where $\epsilon$ is the constant of integration. We refer to \eqref{evo-con-HJ} as the evolution Hamilton-Jacobi equation on the base $M$.  
 Once a solution $W$ is found for the evolution Hamilton-Jacobi equation \eqref{evo-con-HJ}, we can lift it (as a solution to the projected dynamics $\varepsilon^{\mathcal{T}^*W}_H$ on $M$) to a solution to the evolution dynamics $\varepsilon_H$ on the extended cotangent bundle $T^{*}M\times \mathbb{R}$ by means of the first prolongation of $W$.
 
\textbf{Jet Decomposition of Evolution Dynamics.}
  Taking an $n$-dimensional manifold $M$, $T^*M\times \mathbb{R}$ is a $(2n+1)$-dimensional manifold with coordinates $(\mathbf{x},\mathbf{x}^*,z)$, and $J^1\pi^0$ is of dimension $2n+1+(n+1)n$ with the induced local coordinates $(\mathbf{x},\mathbf{x}^*,z,\mathbf{x}^*_\mathbf{x},z_\mathbf{x})$. 
The holonomic lift of a vector field $X=\mathbf{X}\cdot \nabla_\mathbf{x}$ on $M$ determines a generalized vector field 
\begin{equation}
 X ^{hol}(\mathbf{x},\mathbf{x}^*,z,\mathbf{x}^*_\mathbf{x},z_\mathbf{x})=\mathbf{X}\cdot \nabla_\mathbf{x} +\mathbf{x}^*_\mathbf{x}
\mathbf{X} \cdot \nabla_{\mathbf{x}^*}+\mathbf{X}\cdot z_\mathbf{x}\nabla_{z},
 \end{equation}
and the dynamics generated by $X ^{hol}$ is 
\begin{equation} \label{idento-}
 \frac{d \mathbf{x} }{dt} = \mathbf{X} ,\qquad 
 \frac{d \mathbf{x}^*}{dt} =\mathbf{x}^*_\mathbf{x}
\mathbf{X},\qquad  \frac{d z}{dt} = \mathbf{X}\cdot z_\mathbf{x}.
\end{equation}

Let $\varepsilon_H$ be an evolution Hamiltonian dynamics on $T^*M\times \mathbb{R}$. In terms of the local coordinates, the evolution Hamiltonian dynamics determined by a Hamiltonian function $H$ is given in Equation \eqref{evo-dyn}, and its projection $\varepsilon^{\mathcal{T}^*W}_H$ in \eqref{varep-H-proj}, obtained by the first jet prolongation of a real-valued function $W$ on $M$, is
 \begin{equation}
	\varepsilon^{\mathcal{T}^*W}_H=H_{\mathbf{x}^*}
	\Big\vert_{\mathbf{x}^*=W_\mathbf{x},~
z=W	}\cdot \nabla_\mathbf{x}.
\end{equation}
Then, the holonomic lift of $\varepsilon^{\mathcal{T}^*W}_H$ becomes
  \begin{equation}
  \begin{split}
 & (\varepsilon^{\mathcal{T}^*W}_H)^{hol}(\mathbf{x},\mathbf{x}^*,z,\mathbf{x}^*_\mathbf{x},z_\mathbf{x})\\&\hspace{2cm}=H_{\mathbf{x}^*}
	\Big\vert_{\mathbf{x}^*=W_\mathbf{x},~
z=W	}\cdot \nabla_\mathbf{x} +\mathbf{x}^*_\mathbf{x}
H_{\mathbf{x}^*}
	\Big\vert_{\mathbf{x}^*=W_\mathbf{x},~
z=W	} \cdot \nabla_{\mathbf{x}^*}+H_{\mathbf{x}^*}
	\Big\vert_{\mathbf{x}^*=W_\mathbf{x},~
z=W	}\cdot z_\mathbf{x}\nabla_{z}.
\end{split}
   \end{equation}
The first jet prolongation of $\mathcal{T}^*W$ is
   \begin{equation}
   J^1\mathcal{T}^*W:M\longrightarrow J^1\pi^0,\qquad \mathbf{x}\mapsto (\mathbf{x},W_\mathbf{x},W,W_\mathbf{xx},W_\mathbf{x}).
      \end{equation}
Let the holonomic part of the restricted vector field $\varepsilon_H\big\vert_{im \mathcal{T}^*W}$ be denoted by $\mathfrak{H}\varepsilon_H$. 
The projected vector field $\varepsilon^{\mathcal{T}^*W}_H$ and the holonomic part of the evolution Hamiltonian vector field commute in the sense of the following diagram,
\begin{equation}\label{Ham-GENERIC-3}
\xymatrix{J^1\pi^0  \ar[rrr]^{\mathfrak{H}\varepsilon_H}& &
&T(T^*M\times \mathbb{R})   \\ & & &\\
M \ar[uu]^{J^1\mathcal{T}^*W} \ar[rrr]^{\varepsilon^{\mathcal{T}^*W}_H}& & & TM \ar[uu]_{T\mathcal{T}^*W}}
\end{equation}
which also means that 
\begin{equation}
\mathfrak{H}\varepsilon_H \circ J^1\mathcal{T}^*W = T\mathcal{T}^*W\circ \varepsilon^{\mathcal{T}^*W}_H.
\end{equation}
Consequently, the vertical representative 
of the evolution vector field is computed to be
\begin{equation}
\begin{split}
&\mathfrak{V}\varepsilon_H(\mathbf{x},\mathbf{x}^*,z,\mathbf{x}^*_\mathbf{x},z_\mathbf{x})=\Big(-H_\mathbf{x}\Big\vert_{\mathbf{x}^*=W_\mathbf{x},~
z=W	} -{\mathbf{x}^*}H_z\Big\vert_{\mathbf{x}^*=W_\mathbf{x},~
z=W	}  - \mathbf{x}^*_\mathbf{x}
H_{\mathbf{x}^*}
	\Big\vert_{\mathbf{x}^*=W_\mathbf{x},~
z=W	}\Big) \cdot \nabla_{\mathbf{x}^*}\\&\hspace{4cm}+\left({\mathbf{x}^*}\cdot H_{\mathbf{x}^*}\Big\vert_{\mathbf{x}^*=W_\mathbf{x},~
z=W	}-H_{\mathbf{x}^*}
	\Big\vert_{\mathbf{x}^*=W_\mathbf{x},~
z=W	}\cdot z_\mathbf{x}\right)\nabla_{z}.
\end{split}
  \end{equation}  
The evolution geometric Hamilton-Jacobi theorem under the holonomic-vertical decomposition follows.

\begin{proposition} \label{HJT-vert-evo}
For a smooth function $W=W(\xx)$ on $M$, the following conditions are equivalent:
\begin{enumerate}
\item The following is satisfied 
\begin{equation}\label{HJT-Con-1-eq-2}
\mathfrak{H}\varepsilon_H \circ J^1\mathcal{T}^*W  =\varepsilon_H\circ \mathcal{T}^*W,
\end{equation}
where $T\mathcal{T}^*W:TM\mapsto T(T^{*}M\times \mathbb{R})$ is tangent mapping of the prolongation $\mathcal{T}^*W$, whereas $\mathfrak{H}\varepsilon_H$ is the holonomic part.
\item The vertical representative is vanishing that is $\mathfrak{V}\varepsilon_H\circ T\mathcal{T}^*W=0$.
\item The equation is fulfilled that $
d( H\circ \mathcal{T}^*W  )=0.$ 
\end{enumerate}
\end{proposition}

\subsection{Pure Gradient Flow in Evolution Hamiltonian Form}
In Subsection \ref{Grad-Sec}, we have seen a pure dissipative flow \eqref{grad} on a manifold $M$ as a projection of a Hamiltonian flow on $T^*M$. It has also been shown that the lift of solutions of the dissipative flow to the Hamiltonian motion is possible if the entropy is a solution of the Hamilton-Jacobi equation. In the current Section, we show that this relation is also valid in the case of evolution Hamiltonian flow on the extended cotangent bundle $T^*M\times \mathbb{R}$, equipped with the Darboux' coordinates $(\mathbf{x},\mathbf{x}^*,z)$. 

Consider a dissipation potential $\Xi=\Xi(\mathbf{x},\mathbf{x}^*)$ as a function on the extended cotangent bundle $T^*M\times \mathbb{R}$, but independent of the extension coordinate $z$. The evolution Hamiltonian vector field \eqref{evo-dyn} then simplifies to
 \begin{equation}\label{e-H-exp}
	\varepsilon_\Xi=\Xi_{\mathbf{x}^*} \cdot \nabla_\mathbf{x}
-\Xi_\mathbf{x}\cdot \nabla_{\mathbf{x}^*} 
+(\mathbf{x}^*\cdot \Xi_{\mathbf{x}^*})\nabla_z.
 \end{equation} 
Entropy $S$ is a real-valued function on $M$, and its first jet prolongation $\mathcal{T}^*S(\mathbf{x})=(\mathbf{x},S_\mathbf{x},S(\mathbf{x}))$ has values in the extended cotangent bundle $T^*M\times \mathbb{R}$. 

The following diagram summarizes our further steps,
 \begin{equation}\label{Evo-HJ-4}
\xymatrix{  T^{*}M\times \mathbb{R}
\ar[dd]^{\pi^0_M} \ar[rrr]^{\varepsilon_\Xi}&   & &T(T^{*}M\times \mathbb{R})\ar[dd]_{T\pi^0_M}\\
  &  & &\\
M  \ar@/^2pc/[uu]^{\mathcal{T}^*S}\ar[rrr]^{\varepsilon^{\mathcal{T}^*S}_\Xi}&  & & TM \ar@/_2pc/[uu]_{T\mathcal{T}^*S}}
\end{equation}
The projected vector field $\varepsilon^{\mathcal{T}^*S}_\Xi$ on $M$ is
\begin{equation}
\varepsilon^{\mathcal{T}^*S}_\Xi := T\pi^0_M \circ \varepsilon_\Xi \circ \mathcal{T}^*\Phi,
\end{equation}
and the projected motion becomes
\begin{equation}
\varepsilon^{\mathcal{T}^*S}_\Xi = \Xi_{\mathbf{x}^*}
\big\vert_{\mathbf{x}^*=S_\mathbf{x}}\cdot \nabla_\mathbf{x},
\end{equation}
which represents the dissipation flow \eqref{grad}. 

Let us denote the tangent lift of the first jet prolongation of $S$ by $T\mathcal{T}^*S$. The composition of the reduced dynamics $\varepsilon^{\mathcal{T}^*S}_\Xi$ with $T\mathcal{T}^*S$, referring to Diagram \ref{Evo-HJ-4}, starts 
at the left bottom node, goes right, and finally up to the right top node, that is
\begin{equation} \label{T-1-}
T\mathcal{T}^*S\circ \varepsilon^{\mathcal{T}^*S}_\Xi=\Xi_{\mathbf{x}^*}
\big\vert_{\mathbf{x}^*=S_\mathbf{x}}\cdot \nabla_\mathbf{x}+ S_{\mathbf{x}\mathbf{x}}\Xi_{\mathbf{x}^*}
\big\vert_{\mathbf{x}^*=S_\mathbf{x}}\cdot 
\nabla_{\mathbf{x}^*}
+\mathbf{x}^*\cdot \Xi_{\mathbf{x}^*}
\big\vert_{\mathbf{x}^*=S_\mathbf{x}}  \nabla_z,
\end{equation}
where $S_{\mathbf{x}\mathbf{x}}$ is the Hessian matrix of the entropy. 

If, moreover, the dissipation potential is independent of the base component $\mathbf{x}$, then its evolution becomes
\begin{equation} 
\dot{\Xi}= \left\langle d\Xi, T \mathcal{T}^*\Phi  \circ \varepsilon^{\mathcal{T}^*\Phi}_\Xi (x)  \right\rangle =S_{\mathbf{x}\mathbf{x}}\Xi_{\mathbf{x}^*}
\big\vert_{\mathbf{x}^*=S_\mathbf{x}}\cdot \Xi_{\mathbf{x}^*}
\big\vert_{\mathbf{x}^*=S_\mathbf{x}}  \leq 0,
\end{equation}
which corresponds to the principle of least dissipation \cite{prigogine-tip}. 

Finally, the compatibility of the lift of integral curves of the dissipative flow on $M$ to the evolution Hamiltonian flow on $T^*M\times \mathbb{R}$ is expressed in the following variant of the Hamilton-Jacobi Theorem \ref{HJT-Evo-Con-1}.
\begin{proposition}
The evolution Hamiltonian dynamics $\varepsilon_{\Xi}$ in
\eqref{e-H-exp} and the dissipative dynamics $
\varepsilon_{\Xi}^{dS} $ in \eqref{grad} are related by
\begin{equation}
  T\mathcal{T}^*S\circ \varepsilon^{\mathcal{T}^*S}_\Xi = \varepsilon_\Xi \circ \mathcal{T}^*S \label{eq.compat-GENERIC-grad}
\end{equation}
if and only if the entropy is a solution of the evolution Hamilton-Jacobi equation, that is
\begin{equation}
\Xi(\mathbf{x} ,S_\mathbf{x},S(\mathbf{x}))=\epsilon
\end{equation}
for a constant $\epsilon$. 
\end{proposition}  
  
\textbf{Euler-Lagrange Formulation.}
We assume that the dissipation potential $\Xi$ is a convex function, hence we can apply the Legendre transformation and its inverse to it. This gives a Lagrangian function, which we denote by $\Xi^*(\mathbf{x},\dot{\mathbf{x}})$, defined on $TM\times \mathbb{R}$. Here, the transformation is defined as
\begin{equation}
\mathbb{F}\Xi:T^*M\times \mathbb{R}\longrightarrow TM\times \mathbb{R},\qquad (\mathbf{x},\mathbf{x}^*,z) \mapsto (\mathbf{x},\Xi_{\mathbf{x}^*},z),
\end{equation}
that is $\dot{\mathbf{x}}=\Xi_{\mathbf{x}^*}$ in terms of the induced coordinates. 
Since $\Xi$ is independent of $z$, $\Xi^*$ is independent of $z$ and one can determine the evolution Herglotz equations,
\begin{equation}\label{grad-Herglotz}
\frac{d\mathbf{x}}{dt}=\dot{\mathbf{x}},\qquad \Xi^*_\mathbf{x}- \frac{d}{dt}\Xi^*_{\dot{\mathbf{x}}}
= 0, \qquad  \frac{dz}{dt} = \Xi^*(\mathbf{x},\dot{\mathbf{x}}),
\end{equation} 
which are equivalent to the Euler-Lagrange equations \cite{Herglotz}. 

\subsection{GENERIC as an Evolution Hamiltonian Dynamics}\label{sec.Gen.Evo}

Subsection \ref{sectiongeneric} contains a geometric formulation of GENERIC. In the current Section, we extend the formulation to the contact framework, which permits us to add the principle of least dissipation into the system of equations on the lifted level. The contactization is carried out on the extended cotangent bundle $T^*M\times \mathbb{R}$. 

Similarly as in Subsection \ref{Con-Man-Sec}, there are two alternative Hamiltonian formulations on the extended cotangent bundle. One is the contact Hamiltonian flow $X_H$, dissipating the Hamiltonian function, the other is the evolution Hamiltonian flow $\varepsilon_H$ preserving the Hamiltonian. Here, we prefer the latter. 

\textbf{Evolution GENERIC Flow.} First, recall the Hamiltonian function $\Psi$ given in \eqref{Psi}, defined on the cotangent bundle $T^*M$. Due to the canonical inclusion of the cotangent bundle $T^*M$ into the extended cotangent bundle $T^*M\times \mathbb{R}$, $\Psi$ is also a function on $T^*M\times \mathbb{R}$. In the Darboux' coordinates $(\mathbf{x},\mathbf{x}^*,z)$ on $T^*M\times \mathbb{R}$, $\Psi$ turns out to be independent of the fiber variable $z$. Then,  the evolution Hamiltonian flow (Equation \eqref{evo-dyn}) generated by $\Psi$ is
\begin{equation}\label{e-H-exp-2} 
 \varepsilon_\Psi(\mathbf{x},\mathbf{x}^*)=\left( \mathbb{L}\Phi_\mathbf{x}-\Xi_{\mathbf{x}^*}\right)\cdot \nabla_{\mathbf{x}}
+(\Xi_{\mathbf{x}}-\nabla_{\mathbf{x}}(\mathbf{x}^*\cdot \mathbb{L}\Phi_\mathbf{x}))\cdot\nabla_{\mathbf{x}^*}
+\mathbf{x}^*\cdot \left(\frac{1}{e^*}\mathbb{L}\Phi_\mathbf{x}-\Xi_{\mathbf{x}^*}\right) \nabla_z,
\end{equation} 
and the corresponding evolution equations become
\begin{equation}\label{e-H-exp-em}
\frac{d \mathbf{x}}{dt}= \mathbb{L}\Phi_\mathbf{x}-\Xi_{\mathbf{x}^*},   \qquad \frac{d \mathbf{x}^*}{dt} = \Xi_{\mathbf{x}}-\nabla_{\mathbf{x}}(\mathbf{x}^*\cdot \mathbb{L}\Phi_\mathbf{x})\frac{1}{e^*},\qquad \frac{d z}{dt}=\mathbf{x}^*\cdot \left(\frac{1}{e^*}\mathbb{L}\Phi_\mathbf{x}-\Xi_{\mathbf{x}^*}\right).
\end{equation}
We refer to this dynamics as the evolution GENERIC flow.

Note that if we used the contact Hamiltonian vector field $X_\Psi$ (Equation \eqref{con-dyn}) instead of the evolution Hamiltonian vector field $\varepsilon_\Psi$, an extra $-\Psi$ term for the basis ${\partial}
/{\partial z}$ would appear. 

A thermodynamical potential $\Phi$ is a function on the base manifold $M$, and its first prolongation to the extended cotangent bundle as a section is
\begin{equation}\label{T*Phi}
\mathcal{T}^*\Phi:M \longrightarrow T^{*}M\times \mathbb{R},\qquad (x)\mapsto (\mathbf{x},\Phi_\mathbf{x}(\mathbf{x}),\Phi(\mathbf{x})).
\end{equation}
Then, a projected vector field $\varepsilon^{\mathcal{T}^*\Phi}_\Psi $ on $M$ is defined as
 \begin{equation}
\varepsilon^{\mathcal{T}^*\Phi}_\Psi (x):= T\pi^0_M \circ \varepsilon_\Psi \circ \mathcal{T}^*\Phi,
 \end{equation}
or, in coordinates, as
\begin{equation}\label{GENERIC-evolution}
\varepsilon^{\mathcal{T}^*\Phi}_\Psi (\mathbf{x}) = (\mathbb{L}\Phi_\mathbf{x}-\Xi_{\mathbf{x}^*}\Big\vert_{\mathbf{x}^*= \Phi_{\mathbf{x}}})\cdot \nabla_{\mathbf{x}}.
 \end{equation}
The dynamics on the base manifold generated by the projected vector field is the GENERIC flow \eqref{Generic}. 
 
\textbf{Thermodynamical Potential as a Lyapunov Function.}
The composition of the projected vector field $ \varepsilon^{\mathcal{T}^*\Phi}_\Psi$ and the tangent lift $T\mathcal{T}^*\Phi$ of the first prolongation  $\mathcal{T}^*\Phi$ reads
 \begin{equation} \label{T-1}
 T \mathcal{T}^*\Phi  \circ \varepsilon^{\mathcal{T}^*\Phi}_\Psi (\mathbf{x})  =\left(\mathbb{L}\Phi_\mathbf{x}-\Xi_{\mathbf{x}^*}\Big\vert_{\mathbf{x}^*= \Phi_{\mathbf{x}}}\right)\cdot \nabla_{\mathbf{x}} + \Phi_{\mathbf{x}\mathbf{x}} \left(\mathbb{L}\Phi_\mathbf{x}-\Xi_{\mathbf{x}^*}\Big\vert_{\mathbf{x}^*= \Phi_{\mathbf{x}}}\right)\cdot \nabla_{\mathbf{x}^*}
+
 \mathbf{x}^* \cdot \Xi_{\mathbf{x}^*}\Big\vert_{\mathbf{x}^*= \Phi_{\mathbf{x}}}\nabla_{z},
\end{equation}
which corresponds to evolution equations
\begin{equation}\label{dyn-eq-evo}
\frac{d \mathbf{x}}{dt}=\mathbb{L}\Phi_\mathbf{x}-\Xi_{\mathbf{x}^*},   \qquad \frac{d \mathbf{x}^*}{dt} = \Phi_{\mathbf{x}\mathbf{x}} \left(\mathbb{L}\Phi_\mathbf{x}-\Xi_{\mathbf{x}^*}\Big\vert_{\mathbf{x}^*= \Phi_{\mathbf{x}}}\right),\qquad \frac{d z}{dt}=\mathbf{x}^*\cdot (\mathbb{L}\Phi_\mathbf{x}-\Xi_{\mathbf{x}^*}).
\end{equation}
The first system in \eqref{dyn-eq-evo} is the GENERIC flow on $M$, the second system is the rate GENERIC flow, and the third equality in \eqref{dyn-eq-evo} is always negative, 
  \begin{equation}\label{Phi-dot}
  \dot{\Phi}=
- \left( \mathbf{x}^* \cdot \Xi_{\mathbf{x}^*}\right)\Big\vert_{\mathbf{x}^*= \Phi_{\mathbf{x}}}=  -  \Phi_{\mathbf{x}}\cdot  \Xi_{\mathbf{x}^*}\Big\vert_{\mathbf{x}^*= \Phi_{\mathbf{x}}}\leq 0,
  \end{equation}
if the dissipation potential $\Xi$ is convex with respect to the dual coordinates, which expresses the second law of thermodynamics. In other words, Equations \eqref{dyn-eq-evo} combing GENERIC, rate GENERIC, and a direct manifestation of the second law of thermodynamics.

\textbf{GENERIC Flow and the Evolution Hamiltonian Flow.} Let us now discuss the relationship between the evolution Hamiltonian dynamics $\varepsilon_{\Psi}$ on $T^*M\times \mathbb{R}$, given in \eqref{e-H-exp-2}, and GENERIC flow $\varepsilon^{\mathcal{T}^*\Phi}_\Psi$, given in \eqref{GENERIC-evolution}.
First, we consider the composition of the evolution Hamiltonian flow $\varepsilon_H$ and the prolongation $\mathcal{T}^*\Phi$,
  \begin{equation}\label{T-2}
  \begin{split} 
&\varepsilon_\Psi \circ \mathcal{T}^*\Phi (\mathbf{x})=\varepsilon_\Psi(\mathbf{x},\Phi_\mathbf{x}(\mathbf{x}),\Phi(\mathbf{x}))\\&\qquad = \left(\mathbb{L}\Phi_\mathbf{x}-\Xi_{\mathbf{x}^*}\Big\vert_{\mathbf{x}^*= \Phi_{\mathbf{x}}}\right)\cdot \nabla_{\mathbf{x}} 
+\Xi_{\mathbf{x}}\Big\vert_{\mathbf{x}^*=\Phi_{\mathbf{x}}} \cdot \nabla_{\mathbf{x}^*} -\mathbf{x}^* \cdot  \Xi_{\mathbf{x}^*} \Big \vert _{\mathbf{x}^{\ast }=\Phi _{\mathbf{x}}}\nabla_{z}.
 \end{split}
  \end{equation}

Second, the second law of thermodynamics can be obtained by lifting the exterior derivative $d\Phi$ of the thermodynamical potential to the extended cotangent manifold $T^*M \times \mathbb{R}$. This lift $(d\Phi)^h$ (actually a one-form) is then contracted with vector field \eqref{T-2}, which gives
  \begin{equation}
  \dot{\Phi}= \left\langle \varepsilon_\Psi \circ \mathcal{T}^*\Phi , (d\Phi)^h\right \rangle(\mathbf{x})  =\left(\frac{1}{e^*}\mathbb{L}\Phi_\mathbf{x}-\Xi_{\mathbf{x}^*}\Big\vert_{\mathbf{x}^*= \Phi_{\mathbf{x}}}\right)\cdot \Phi_{\mathbf{x}} =- \Xi_{\mathbf{x}^*}\Big\vert_{\mathbf{x}^*= \Phi_{\mathbf{x}}}\cdot \Phi_{\mathbf{x}},
    \end{equation}
where $\varepsilon_{\Psi} $ is the evolution vector field from Equation \eqref{e-H-exp-2}. 

The variant of the Hamilton-Jacobi theorem \ref{HJT-Evo-Con-1} for evolution Hamiltonian dynamics in the present context follows.
  \begin{proposition}\label{Prop-Evo}
  Dynamics given in \eqref{T-1} and \eqref{T-2} are equivalent, that is
  \begin{equation} 
\varepsilon_\Psi \circ \mathcal{T}^*\Phi (\xx)=T \mathcal{T}^*\Phi  \circ \varepsilon^{\mathcal{T}^*\Phi}_\Psi (\xx),
  \end{equation}
  if and only if the thermodynamic potential $\Phi$ is a solution of the Hamilton-Jacobi problem for the (evolution) Hamiltonian function $\Psi$, 
  \begin{equation} 
  \Psi(\xx,\Phi_{\xx}(\xx),\Phi(\xx))=-\Xi(\xx, \Phi_{\xx})= \epsilon.
    \end{equation}
    \end{proposition}
In this case, one can lift a solution of GENERIC on the base manifold $M$ to a solution of the evolution GENERIC flow \eqref{dyn-eq-evo} on the extended cotangent manifold $T^*M\times \mathbb{R}$.

\textbf{A Simple Example: A particle with friction.}
Let us now demonstrate this evolution Hamiltonian GENERIC on a simple example, a particle with friction. The state variables are position, momentum, and entropy, $\mathbf{x} = (q,p,s)$, and the thermodynamic potential is $\Phi(\mathbf{x}) = s - e^* e(q,p,s)$. The Poisson bivector and dissipation potential are 
\begin{equation}
\mathbb{L} = \begin{pmatrix} 0 & 1 & 0\\ -1 & 0 & 0\\ 0 & 0 & 0\end{pmatrix},\qquad  \Xi(\mathbf{x},\mathbf{x}^*) = \frac{1}{2}\zeta \left(p^* -\frac{e_p}{e_s} s^*\right)^2, 
\end{equation}
respectively, see \cite{JSP2020}. Then the vector field $\epsilon_\Psi\circ\mathcal{T}^*\Phi$ generates evolution equations
\begin{equation}
\frac{d}{dt}\begin{pmatrix}q\\p\\s\end{pmatrix}
= 
\begin{pmatrix}
 0 & 1 & 0\\ -1 & 0 & 0\\ 0 & 0 & 0
\end{pmatrix}
\cdot
\begin{pmatrix}
e_q\\
e_p\\
e_s
\end{pmatrix}
- 
\zeta 
\begin{pmatrix}
0\\
p^*-\frac{e_p}{e_s}s^*\\
-\left(p^*-\frac{e_p}{e_s}s^*\right)\frac{e_p}{e_s}
\end{pmatrix}\Big|_{\mathbf{x}^*=\Phi_{\mathbf{x}}}
=
\begin{pmatrix}
e_p\\
-e_q\\
0
\end{pmatrix}
-\zeta
\begin{pmatrix}
0\\
\frac{e_p}{e_s}\\
-\left(\frac{e_p}{e_s}\right)^2
\end{pmatrix}.
\end{equation}
Taking energy as the sum of kinetic, potential, and internal, $e=p^2/2m+V(q)+e_{int}(s)$, and choosing the friction coefficient as $\zeta = \zeta_0   e_s$, where $\zeta_0$ is a positive constant, we obtain the usual equations for a particle that is moving in a potential while experiencing linear friction. The derivative $e_s=T$ is the temperature of the (macroscopic) particle and is always positive. The entropy is always growing, which represents the second law of thermodynamics. Moreover, if we choose $\zeta = \zeta_0  e^2_s/ e_p^2 = \zeta_0 mT^2$, then we satisfy Hamilton-Jacobi equation which means that the lifted evolution is equivalent to the original one. This is a generalization of the usual evolution Hamiltonian dynamics \cite{simoes2020contact} because we can prescribe any friction coefficient, due to the presence of the dissipation potential.
 
\textbf{Euler-Lagrange Formulation.}
There are two convex functions on the extended cotangent bundle, the dissipation potential $\Xi$ and its extension $\Psi$ with the momentum function. Therefore, we can compute the Legendre transformation of $\Psi$ and arrive at the evolution Herglotz Lagrangian realization of the evolution GENERIC flow.
The Legendre transformation is 
\begin{equation}
\mathbb{F}\Psi:T^*M\times \mathbb{R}\longrightarrow TM\times \mathbb{R},\qquad (\mathbf{x},\mathbf{x}^*,z) \mapsto (\mathbf{x},\Psi_{\mathbf{x}^*},z),
\end{equation}
that is 
\begin{equation}
\dot{\mathbf{x}}=\Psi_{\mathbf{x}^*}=\mathbb{L}\Phi_\mathbf{x}-\Xi_{\mathbf{x}^*}
\end{equation} 
 in terms of the induced coordinates. Denoting the Lagrangian as $\Psi^*$, we arrive at the following system of equations
    \begin{equation} 
\frac{d\mathbf{x}}{dt}=\dot{\mathbf{x}},\qquad \Psi^*_\mathbf{x}- \frac{d}{dt}\Psi^*_{\dot{\mathbf{x}}} = 0, \qquad  \frac{dz}{dt} =\dot{\mathbf{x}}\cdot \Psi^*_{\dot{\mathbf{x}}},
   \end{equation}
which is equivalent to the Euler-Lagrange equations.

\subsection{The Legendrian Submanifolds and Rate-GENERIC Flow}

Consider a $(2n+1)-$dimensional contact manifold equipped with a contact one-form $\eta$. A maximally integrable submanifold of the contact manifold where the contact form vanishes is called Legendrian submanifold, see, for example, \cite{Arnold-book,Bravetti19,esen2021contact,Grmela-contact}. It is possible to see that a Legendrian submanifold is necessarily of dimension $n$. In Darboux' coordinates $(\xx,\xx^*,z)$, consider a partition $A\cup B$ of the set of indices $(1,\dots, n)$ into two disjoint subsets, so that the Darboux' coordinates turn out to be
\begin{equation}
(\xx,\xx^*,z)=(q^a,q^\alpha,p_a,p_\alpha,z)
\end{equation}
 where $a\in A$ and $\alpha\in B$.  For a function $\Phi( {q}^a, {p}_\alpha)$ of $n$ variables $ {q}^a$, $a\in A$ and $ {p}_\alpha$, $\alpha\in B$, the
$2n + 1$ equations 
\begin{equation}\label{Leg-Sub-Loc}
N=\left\{(q^a,q^\alpha,p_a,p_\alpha,z): {q}^\alpha=-\frac{\partial \Phi }{\partial {p}_\alpha},
{p}_a=\frac{\partial \Phi }{\partial  {q}^a}, z=\Phi- {p}_\alpha \frac{\partial \Phi }{\partial  {p}_\alpha}  \right\}
\end{equation}
define a Legendrian submanifold of $M$. The inverse of this assertion is also true \cite{Arnold-book}, every Legendrian submanifold can locally be written in the form \eqref{Leg-Sub-Loc}.

To be more concrete, let us concentrate on the extended cotangent bundle $T^*M\times \mathbb{R}$ where $M$ is $n$-dimensional, considering two extreme decompositions $A\cup B$ of the set of indices $(1,\dots, n)$. For the first extreme case, consider a function $\Phi=\Phi(\xx)$ defined on the base manifold $M$. Its first prolongation $\mathcal{T}^*\Phi$ is a section of the extended cotangent bundle, and the image space is a Legendrian submanifold
\begin{equation}\label{N-Phi}
N_\Phi=\left\{(\xx,\xx^*,z)\in T^*M\times \mathbb{R}: ~ \xx^*=\Phi_\xx(\xx), ~ z=\Phi(\xx) \right\}.
\end{equation}
For the other extreme case, take a function that depends only on the momentum varibales, $\Upsilon=\Upsilon(\xx^*)$. The Legendrian submanifold is then
\begin{equation}\label{N-Upsilon}
N_\Upsilon=\left\{(\xx,\xx^*,z)\in T^*M\times \mathbb{R}: ~ \xx=-\Upsilon_{\xx^*}(\xx^*), ~ z=\Upsilon(\xx^*)-\xx^*\cdot \Upsilon_{\xx^*} (\xx^*) \right\}.
\end{equation}
These two Legendrian submanifolds are related by a Legendre transformation, see \cite{Goto-15}. To see this more explicitly, we first assume that $M$ admits a vector space structure and define a section
\begin{equation}\label{D-Upsilon}
D\Upsilon: M^*\longrightarrow T^*M^*\times \mathbb{R},\qquad \xx^*\mapsto (\xx,\Upsilon_{\xx^*}(\xx^*),
\xx^*\cdot \Upsilon_{\xx^*} (\xx^*)-\Upsilon(\xx^*))
\end{equation}
of the extended cotangent bundle $T^*M^*\times \mathbb{R}$ of the dual space $M^*$. Minus of the image space of $D\Upsilon$ is precisely determining the Legendrian submanifold $N_\Upsilon$ in \eqref{N-Upsilon}. 

Now, we define the Legendre transformation of $\Phi=\Phi(\xx)$ as given in \eqref{Leg-dual}. Under the assumption of the regularity, we also have a function $\Phi^*=\Phi^*(\xx^*)$ defined on $M^*$. 
Then we have the following commutative diagram,
\begin{equation}
\xymatrix{ T^{*}M=M\times M^*\times \mathbb{R} \ar[dd]^{\pi_0} \ar[rrr]^{\hat{\Gamma}}& &
& T^{\ast }M^{\ast }\times \mathbb{R}=M^{\ast }\times M \times \mathbb{R} \ar[dd]_{\pi_{0}} \\ & & &\\
M\ar@/^2pc/[uu]^{\mathcal{T}^*\Phi}\ar[rrr]^{Leg}& & & M^* \ar@/_2pc/[uu]_{D\Phi^*}}
\end{equation}
where $Leg$ is the induced Legendre transformation and $\hat{\Gamma}$ is  
\begin{equation}\label{Leg-dual-}
\hat{\Gamma} (\xx,\xx^*,z)=(\xx^*,\xx,z).
\end{equation}
Note that $\mathcal{T}^*\Phi$ is in the form of Equation \eqref{T*Phi} while $D\Phi^*$ is calculated according to Equation \eqref{D-Upsilon}. 


\textbf{Mrugala Metric.} Any contact manifold $(M,\eta)$, equipped with Darboux' coordinates, admits a semi-Riemanian Mrugala metric \cite{Mrugala}
\begin{equation}
\mathfrak{G}=d\xx\otimes^s d\xx^*+\eta\otimes\eta,
\end{equation}
where the term $d\xx\otimes^s d\xx^*$ is obtained by the symmetrization of the tensor product $d\xx\otimes d\xx^*$. 

If we restrict this metric to the Legendrian submanifolds $N_\Phi$ and $N_\Upsilon$, we arrive at 
\begin{equation}
\mathfrak{G}_\Phi=\Phi_{\xx \xx}d\xx\otimes d\xx,\qquad 
\mathfrak{G}_\Upsilon=\Upsilon_{\xx^*\xx^*}d\xx^*\otimes d\xx^*,
\end{equation}
respectively. These two induced structures are Riemannian metrics on the respective submanifolds. A more detailed analysis can reveal that these Legendrian submanifolds are examples of Hessian manifolds \cite{matsuzoe2013geometric,Shima-Yagi}, see  \cite{Goto-15}.

If the Legendre transformation is applied to the metric $\mathfrak{G}_\Phi$, one arrives at the following Riemanian space $M^*$ equipped with metric tensor
\begin{equation}
\mathfrak{G}_{\Phi^*}=[\Phi^*_{\xx^* \xx^*}]^{-1}d\xx^*\otimes d\xx^*. 
\end{equation}
On this Legendrian submanifold, the pure gradient flow generated by the functions $\Psi$ is 
\begin{equation}
\frac{d\xx^*}{dt}=\mathfrak{G}_{\Phi^*} \Psi_{\xx^*}\big\vert_{\xx=\Phi^*_{\xx^*}}=[\Phi^*_{\xx^* \xx^*}]^{-1} 
\Psi_{\xx^*}\big\vert_{\xx=\Phi^*_{\xx^*}},
\end{equation}
which is precisely equal to the rate GENERIC motion \eqref{Rate-GENE}. 

\subsection{Rate GENERIC as Lift to the Extended Iterated Cotangent Bundle}
Consider the contact manifold $T^{\ast }T^{\ast }M\times  
\mathbb{R} 
$ by adding a trivial line bundle to the iterated cotangent bundle. 
The induced local coordinates for $T^{\ast }T^{\ast }M\times 
\mathbb{R}
$ are $\left( \mathbf{x},\mathbf{x}^{\ast };\mathbf{x}^{\top }\mathbf{,x}%
^{\dag },z\right) $ where $z$ stands for an element in $
\mathbb{R}
$.  
For the Hamiltonian function $H=H\left( \mathbf{x},\mathbf{x}^{\ast };%
\mathbf{x}^{\top }\mathbf{,x}^{\dag },z\right) $,  the evolution vector field
on the contact manifold $T^{\ast }T^{\ast }M\times 
\mathbb{R}
$ is
\begin{equation}
\varepsilon _{H}=H_{\mathbf{x}^{\top }}\cdot \nabla _{\mathbf{x}}+H_{\mathbf{%
x}^{\dag }}\cdot \nabla _{\mathbf{x}^{\ast }}-(H_{\mathbf{x}}+\mathbf{x}%
^{\top }H_{z})\cdot \nabla _{\mathbf{x}^{\top }}-(H_{\mathbf{x}^{\ast }}+%
\mathbf{x}^{\dag }H_{z})\cdot \nabla _{\mathbf{x}^{\dag }}+(\mathbf{x}^{\top
}\cdot H_{\mathbf{x}^{\top }}+\mathbf{x}^{\dag }\cdot H_{\mathbf{x}^{\dag
}})\nabla _{z},
\end{equation}%
and the corresponding evolution equations are 
\begin{equation}
\begin{split}
\frac{d\mathbf{x}}{dt}&=H_{\mathbf{x}^{\top }},\qquad \frac{d\mathbf{x}^{\ast
}}{dt}=H_{\mathbf{x}^{\dag }},\qquad \frac{d\mathbf{x}^{\top }}{dt}=-H_{%
\mathbf{x}}-\mathbf{x}^{\top }H_{z},\qquad \frac{d\mathbf{x}^{\dag }}{dt}%
=-H_{\mathbf{x}^{\ast }}-\mathbf{x}^{\dag }H_{z},\\ \frac{dz}{dt}&=
\mathbf{x}^{\top }\cdot H_{\mathbf{x}^{\top }}+\mathbf{x}^{\dag }\cdot H_{%
\mathbf{x}^{\dag }}.   
\end{split}
\end{equation}

Now we recall the static potential, dynamic
potential and dynamic super potential, given in Equation \eqref{fcts}. As before, $\Omega$ is chosen as the Hamiltonian function, it is independent of
the covector variable $\mathbf{x}^{\top }$ and the fiber variable $z$, and it is a
quadratic function of $\mathbf{x}^{\dag }$. 
The evolution vector field then becomes
\begin{equation*}
\varepsilon _{\Omega }=S_{\mathbf{xx}}\mathbf{x}^{\dag }\cdot \nabla _{%
\mathbf{x}^{\ast }}-\frac{1}{2}\left\langle \mathbf{x}^{\dag },S_{\mathbf{xx}%
}\mathbf{x}^{\dag }\right\rangle _{\mathbf{x}}\cdot \nabla _{\mathbf{x}%
^{\top }}+\left\langle \mathbf{x}^{\dag },S_{\mathbf{xx}}\mathbf{x}^{\dag
}\right\rangle \nabla _{z},
\end{equation*}%
whereas the non-zero dynamical equations are%
\begin{equation*}
\frac{d\mathbf{x}^{\ast }}{dt}=S_{\mathbf{xx}}\mathbf{x}^{\dag },\qquad 
\frac{d\mathbf{x}^{\top }}{dt}=-\frac{1}{2}\left\langle \mathbf{x}^{\dag
},S_{\mathbf{xx}}\mathbf{x}^{\dag }\right\rangle _{\mathbf{x}},\qquad \frac{%
dz}{dt}=\left\langle \mathbf{x}^{\dag },S_{\mathbf{xx}}\mathbf{x}^{\dag
}\right\rangle .
\end{equation*}%

Let us now examine the dynamics when restricted to the image of the
Legendrian submanifold $im\left( \mathcal{T}^{\ast }\Psi \right) $, taking $\Psi$ from Equation \eqref{fcts} on the
contact manifold $T^{\ast }T^{\ast }M\times 
\mathbb{R}
$, determined by the first jet of the dynamic potential $\Psi$. In this case,
we have that%
\begin{equation*}
\mathcal{T}^{\ast }\Psi \left( \mathbf{x},\mathbf{x}^{\ast }\right) =\left( 
\mathbf{x},\mathbf{x}^{\ast };\Psi _{\mathbf{x}},\Psi _{\mathbf{x}^{\ast
}},\Psi \left( \mathbf{x},\mathbf{x}^{\ast }\right) \right) =\left( \mathbf{x%
},\mathbf{x}^{\ast };-\Xi _{\mathbf{x}},-\Xi _{\mathbf{x}^{\ast }},-\Xi
\left( \mathbf{x,x}^{\ast }\right) \right), 
\end{equation*}%
which means that the evolution vector field is 
\begin{equation*}
\varepsilon _{H}\circ \mathcal{T}^{\ast }\Psi =S_{\mathbf{xx}}\Psi _{\mathbf{%
x}^{\ast }}\cdot \nabla _{\mathbf{x}^{\ast }}-\frac{1}{2}\left\langle \Psi _{%
\mathbf{x}^{\ast }},S_{\mathbf{xx}}\Psi _{\mathbf{x}^{\ast }}\right\rangle _{%
\mathbf{x}}\cdot \nabla _{\mathbf{x}^{\top }}+\left\langle \Psi _{\mathbf{x}%
^{\ast }},S_{\mathbf{xx}}\Psi _{\mathbf{x}^{\ast }}\right\rangle \nabla _{z}
\end{equation*}%
and the dynamical equations become
\begin{equation*}
\frac{d\mathbf{x}^{\ast }}{dt}=S_{\mathbf{xx}}\Psi _{\mathbf{x}^{\ast
}},\qquad \frac{d\mathbf{x}^{\top }}{dt}=-\frac{1}{2}\left\langle \Psi _{%
\mathbf{x}^{\ast }},S_{\mathbf{xxx}}\Psi _{\mathbf{x}^{\ast }}\right\rangle
,\qquad \frac{dz}{dt}=\left\langle \Psi _{\mathbf{x}^{\ast }},S_{\mathbf{xx}%
}\Psi _{\mathbf{x}^{\ast }}\right\rangle .
\end{equation*}%

By projecting the vector field to the base manifold $T^{\ast }M$, we have the
projected vector field $\varepsilon _{\Omega}^{\mathcal{T}^{\ast }\Psi }$, see the following diagram:
\begin{equation}
\xymatrix{  T^{*}T^{*}M\times \mathbb{R} \ar[dd]^{\pi^0_{T^{*}M}}
\ar[rrr]^{\varepsilon_\Omega}& & &T(T^{*}T^{*}M\times
\mathbb{R})\ar[dd] \\ & & &\\ T^*M \ar@/^2pc/[uu]^{\mathcal{T}^*\Psi} \ar[rrr]^{\varepsilon _{\Omega}^{\mathcal{T}^{\ast }\Psi }}&
& & TT^*M  }   
\end{equation}
Explicitly, we have that
\begin{equation*}
\varepsilon _{\Omega}^{\mathcal{T}^{\ast }\Psi }:=\pi _{TT^{\ast }M}\circ
\varepsilon _{\Omega}\circ \mathcal{T}^{\ast }\Psi =S_{\mathbf{xx}}\Psi _{\mathbf{%
x}^{\ast }}\cdot \nabla _{\mathbf{x}^{\ast }}
\end{equation*}%
and the dynamics%
\begin{equation*}
\frac{d\mathbf{x}^{\ast }}{dt}=S_{\mathbf{xx}}\Psi _{\mathbf{x}^{\ast }},
\end{equation*}
where $\xx$ is a set of parameters.  

\textbf{With Lagrange Multipliers.} Let us, eventually, add also constraints to the dynamics. When $M$ is a vector space, identities
\begin{equation*}
T^{\ast }M\cong T^{\ast }M^{\ast },\text{ \ \ }\left( \mathbf{x},\mathbf{x}%
^{\ast }\right) \Longleftrightarrow \left( \mathbf{x}^{\ast },\mathbf{x}%
\right)
\end{equation*}%
mean that $\Psi =-\Xi \ $ is also a function on $T^{\ast }M^{\ast }
$. Now we replace $\Psi $ with 
\begin{equation*}
\Psi :Y^{\dag }\rightarrow 
\mathbb{R}
,\text{ \ \ }\left( \mathbf{x},\mathbf{x}^{\ast },\mathbf{y}^{\dag }\right)
\rightarrow -\Xi \left( \mathbf{x,x}^{\ast }\right) +\left\langle \mathbf{y}%
^{\dag },\mathbf{y}\left( \mathbf{x}\right) \right\rangle 
\end{equation*}%
where $Y^{\dag }$ is the total space of the fibration $Y^{\dag }\rightarrow
T^{\ast }M^{\ast }$. 

$\Psi $ turns out to be a Morse family if the rank of
the matrix $\mathbf{y}_{\mathbf{x}}$ is full. In this case, 
the Legendrian submanifold determined by $\Psi $ is
\begin{equation*}
R_{\Psi }=\left\{ \left( \mathbf{x},\mathbf{x}^{\ast },\Psi _{\mathbf{x}%
}\left( \mathbf{x},\mathbf{x}^{\ast },\mathbf{y}^{\dag }\right) ,\Psi _{%
\mathbf{x}^{\ast }}\left( \mathbf{x},\mathbf{x}^{\ast },\mathbf{y}^{\dag
}\right) \right) \in T^{\ast }M:\mathbf{y}\left( \mathbf{x}\right)
=0\right\} 
\end{equation*}%
and the evolutionary vector field restricted to that manifold becomes 
\begin{equation*}
\varepsilon _{\Omega }\big\vert_{R_{\Psi }}=S_{\mathbf{xx}}\Psi _{\mathbf{x}%
^{\ast }}\cdot \nabla _{\mathbf{x}^{\ast }}-\frac{1}{2}\left\langle \Psi _{%
\mathbf{x}^{\ast }},S_{\mathbf{xx}}\Psi _{\mathbf{x}^{\ast }}\right\rangle _{%
\mathbf{x}}\cdot \nabla _{\mathbf{x}^{\top }}+\left\langle \Psi _{\mathbf{x}%
^{\ast }},S_{\mathbf{xx}}\Psi _{\mathbf{x}^{\ast }}\right\rangle \nabla
_{z},\qquad \mathbf{y}\left( \mathbf{x}\right) =0.
\end{equation*}%
The corresponding dynamical equations are%
\begin{equation*}
\frac{d\mathbf{x}^{\ast }}{dt}=S_{\mathbf{xx}}\Psi _{\mathbf{x}^{\ast
}},\qquad \frac{d\mathbf{x}^{\top }}{dt}=-\frac{1}{2}\left\langle \Psi _{%
\mathbf{x}^{\ast }},S_{\mathbf{xxx}}\Psi _{\mathbf{x}^{\ast }}\right\rangle
,\qquad \frac{dz}{dt}=\left\langle \Psi _{\mathbf{x}^{\ast }},S_{\mathbf{xx}%
}\Psi _{\mathbf{x}^{\ast }}\right\rangle ,\qquad \mathbf{y}\left( \mathbf{x}%
\right) =0.
\end{equation*}%

Finally, projecting $\varepsilon _{\Omega }\big\vert_{R_{\Psi }}$ to the base manifold $T^{\ast }M^{\ast }$ with $T\pi^0_{T^{*}M} $, we arrive
at implicit dynamics $T\pi^0_{T^{*}M}(\varepsilon _{\Omega }\big\vert_{R_{\Psi }})$,
\begin{equation*} 
\frac{d\mathbf{x}^{\ast }}{dt}=S_{\mathbf{xx}}\Psi _{\mathbf{x}^{\ast
}},\qquad \mathbf{y}\left( \mathbf{x}\right) =0,
\end{equation*}
where $\xx$ is a set of parameters. 
Relations among the lifted and projected vector fields are summarized in the following diagram:
\begin{equation}
\xymatrix{ \mathbb{R}& Y^{\dag} \ar[dd]^{\tau}\ar[l]_{\Psi}& \ar@(ul,ur)^{R_{\Psi}}  T^{*}T^{*}M\times \mathbb{R} \ar[dd]^{\pi^0_{T^{*}M}}
\ar[rrr]^{\varepsilon_\Omega}& & &T(T^{*}T^{*}M\times
\mathbb{R})\ar[dd] \\ & & &\\ &
M  \ar@{=}[r]& T^*M   \ar[rrr]^{\varepsilon _{\Omega}^{\mathcal{T}^{\ast }\Psi }}&
& & TT^*M  \ar@(dl,dr)_{T\pi^0_{T^{*}M}(\varepsilon _{\Omega }\big\vert_{R_{\Psi }})} }  \label{Evo-HJ-1}
\end{equation} 

\section{Conclusion}

In this work, we have found geometric settings for GENERIC and rate GENERIC dynamics, as well as relations among them. The main idea is to lift both the Hamiltonian and the gradient parts of GENERIC on a manifold $M$ to the cotangent bundle $T^*M$, where they become realizations of the same geometry. Alternatively, they can be lifted to evolutionary vector fields on the extended cotangent bundle $T^*M\times \mathbb{R}$ (contact geometry), where also the second law of thermodynamics becomes explicitly part of the governing equations.

In Section \ref{3}, we have presented GENERIC as a projection of a Hamiltonian flow on the symplectic bundle $T^*M$, see Equation \eqref{Generic-pre}. The lift of GENERIC to the Hamiltonian dynamics on the cotangent bundle level is equivalent with the original GENERIC if and only if the thermodynamic potential solves a stationary Hamilton-Jacobi equation, see Proposition \ref{Prop-Ham}.  
Moreover, the holonomic lift of GENERIC flow to the cotangent bundle leads to the splitting to the holonomic and vertical parts, and the vertical part can be exploited to for a geometric reduction of the equations. The holonomic lift also determines the rate GENERIC on the momentum variables \eqref{r-GENERIC}. This dynamics turns out to be GENERIC if the rate constitutive equations \eqref{GG} are determined. 

Section \ref{4} casts GENERIC, rate GENERIC, and the second law of thermodynamics into a single system of equations \eqref{dyn-eq-evo}, which we call the evolution GENERIC flow. This is achieved using evolution Hamiltonian dynamics on the extended cotangent bundle $T^*M\times \mathbb{R}$, which admits a contact structure. Proposition \ref{Prop-Evo} then tells that the lift of the GENERIC flow to the evolution GENERIC is equivalent with the original GENERIC if and only if the thermodynamic potential solves a Hamilton-Jacobi equation for evolution Hamiltonian dynamics. Moreover, the evolution GENERIC flow facilitates the formulation of dissipative systems within contact geometry.

The thermodynamic perspective of these geometric results is covered in a follow-up paper \cite{OMM-2}.

\section{Acknowledgments}
MP was supported by Czech Science Foundation, project no. 20-22092S, and by Charles University Research program No. UNCE/SCI/023.


\end{document}